\documentclass[onecolumn,tighten,times]{aastex63}
\usepackage{graphicx}
\usepackage{subfigure}
\usepackage{amsmath}
\usepackage{natbib}
\usepackage{amssymb}
\usepackage{sidecap}
\usepackage{cancel}

\def\prsone{FRB\,20121102}
\def\prstwo{FRB\,20190520B}

\newcommand{\be}{\begin{equation}}
\newcommand{\ee}{\end{equation}}

\begin{document}

\title{Radio Nebul\ae\ from Hyper-Accreting X-ray Binaries as Common Envelope Precursors and Persistent Counterparts of Fast Radio Bursts}

\author[0000-0002-5519-9550]{Navin Sridhar}
\affiliation{Department of Astronomy, Columbia University, New York, NY 10027, USA}
\affiliation{Theoretical High Energy Astrophysics (THEA) Group, Columbia University, New York, New York 10027, USA}
\author[0000-0002-4670-7509]{Brian D. Metzger}
\affil{Department of Physics, Columbia University, New York, NY 10027, USA}
\affiliation{Theoretical High Energy Astrophysics (THEA) Group, Columbia University, New York, New York 10027, USA}
\affil{Center for Computational Astrophysics, Flatiron Institute, 162, 5th Ave, New York, NY 10010, USA}

\begin{abstract}
Roche lobe overflow from a donor star onto a black hole or neutron star binary companion can evolve to a phase of unstable runaway mass-transfer, lasting as short as hundreds of orbits ($\lesssim 10^{2}$ yr for a giant donor), and eventually culminating in a common envelope event.  The highly super-Eddington accretion rates achieved during this brief phase ($\dot{M} \gtrsim 10^{5}\dot{M}_{\rm Edd})$ are accompanied by intense mass-loss in disk winds, analogous but even more extreme than ultra-luminous X-ray (ULX) sources in the nearby universe.  Also in analogy with observed ULX, this expanding outflow will inflate an energetic `bubble' of plasma into the circumbinary medium.  Embedded within this bubble is a nebula of relativistic electrons heated at the termination shock of the faster $v \gtrsim 0.1$\,c wind/jet from the inner accretion flow.  We present a time-dependent, one-zone model for the synchrotron radio emission and other observable properties of such ULX ``hyper-nebul\ae''.  If ULX jets are sources of repeating fast radio bursts (FRB), as recently proposed, such hyper-nebul\ae\ could generate persistent radio emission and contribute large and time-variable rotation measure to the bursts, consistent with those seen from \prsone~and \prstwo.  ULX hyper-nebul\ae\ can be discovered independent of an FRB association in radio surveys such as VLASS, as off-nuclear point-sources whose fluxes can evolve significantly on timescales as short as years, possibly presaging energetic transients from common envelope mergers.
\end{abstract}


\section{Introduction}

A key stage in the ``field'' channel for forming tight neutron star (NS) and black hole (BH) binaries which merge due to gravitational waves is the common envelope interaction between a massive evolved donor star and an orbiting NS or BH companion (e.g., \citealt{Belczynski+02,Dominik+12,Vigna-Gomez+18,Law-Smith+20}).  Unfortunately, the process by which mass transfer via Roche Lobe overflow (RLOF) becomes unstable, leading to the immersion and inspiral of the BH/NS through the envelope of the donor companion star (e.g., \citealt{MacLeod&Loeb20,Marchant+21}), remains poorly understood.  For similar reasons, the final binary separations which result once the common envelope is removed also remain poorly constrained by theory (e.g., \citealt{Ivanova+13}), thereby limiting our ability to make accurate predictions for gravitational wave source populations (e.g., \citealt{Broekgaarden+21,vanSon+22}).  

One approach to addressing these open questions is by directly observing common envelope events in real time.  The ejection of envelope material from one or both stars in a stellar merger event is accompanied by a $\sim$month-long optical/infrared transient known as a ``luminous red nova'' (LRN; e.g., \citealt{Tylenda+11}).  However, the low luminosities $L \lesssim 10^{41}$\,erg\,s$^{-1}$ of LRN, powered by the shock-heated ejecta and energy released by hydrogen recombination (e.g., \citealt{Soker&Tylenda06,Matsumoto&Metzger_22}), limit their detection to the nearby universe (within the Milky Way or nearby galaxies), compared for instance to supernovae.  On the other hand, if the donor star envelope is not removed and the inspiraling NS/BH is driven to merge with the donor's helium core, the tidal disruption and hyper-accretion of the dense core onto the NS/BH may power a more energetic explosion (e.g., \citealt{Chevalier_12,Soker+19,Schroder+20,Dong+21,Metzger22}).  Such ``failed'' common envelope events may be connected to the nascent class of engine-powered ``fast blue optical transients'' (FBOT; e.g., \citealt{Drout+14,Margutti+19,Ho+21b}), which though much rarer than LRNe, can be detected to far greater distances.

This paper considers another source of electromagnetic precursor emission of putative gravitational wave progenitor binaries, from the binary mass-transfer phase which precedes the common envelope and any concomitant LRN or FBOT transient.   Observations of stellar mergers (such as the well-studied event V1309 Sco; \citealt{Tylenda+11,Pejcha14,Mason&Shore22}) indicate that the process of dynamically unstable mass-transfer is not instantaneous, but rather takes place over many tens or hundreds of binary orbital periods (e.g., \citealt{Pejcha+17,MacLeod&Loeb20}).  Applying similar considerations to massive star binaries with BH/NS accretors, the mass-transfer rate during this pre-dynamical phase can exceed the Eddington rate by many orders of magnitudes.  In some systems, a less extreme, but still super-Eddington, thermal-timescale mass-transfer phase will precede the dynamical one (e.g., \citealt{Pavlovskii+17,Marchant+21,Klencki+21}).  Thus, in the millenia to years prior to the onset of the common envelope, the accretion flow onto the BH/NS is similar to, if not more extreme than, those believed to characterize the `ultra-luminous X-ray' (ULX) sources in nearby galaxies (e.g., \citealt{Feng&Soria11,Kaaret+17}).  

Super-Eddington accretion is predicted on theoretical ground to be accompanied by powerful outflows of mass and energy in the form of wide-angle disk winds and collimated relativistic jets (e.g., \citealt{Blandford&Begelman99,Sadowski&Narayan15,Sadowski&Narayan16}), and indeed many ULX systems exhibit evidence for such outflows (e.g., \citealt{Begelman+06,Poutanen+07,Kawashima+12,Middleton+15,Narayan+17}).  Large scale ($>100$\,pc) ionized nebul\ae\ are observed around some ULXs: so-called ULX `bubbles' (e.g., \citealt{Pakull&Mirioni02,Roberts+03,Pakull+06,Kaaret&Feng09})\footnote{Such jet-inflated bubbles---although much smaller in size---have also been observed surrounding non-ULXs like Cyg X-1 \citep{Russell+07}.}.  The size and expansion rates of known ULX nebul\ae\ point to typical source ages of millions of years (e.g., \citealt{Kaaret+17}). An energy input of $\sim 10^{52}-10^{53}\,{\rm erg}$ is required to explain the size, age, and luminosities of these nebul\ae\ (e.g., \citealt{Roberts+03,Pakull+10,Soria+21}), well beyond that supplied by typical supernovae \citep{Asvarov_06} but consistent with that released by the accretion of several solar masses onto a BH if $\sim 10\%$ of the liberated gravitational energy goes into outflow kinetic energy.  ULX nebul\ae\ are observed to emit radio synchrotron emission and optical lines characteristic of shock-ionized gases \citep{Miller+05, Soria+06, Lang+07}, supporting their power sources being continuous disk and jetted outflows from the central accretor.

In the present paper, we develop a model for ULX nebul\ae\ which we extend to much shorter-lived binary outflows that immediately precede common envelope events.  We dub these hypothesized sources ``ULX hyper-nebul\ae'' due to their much higher outflow powers compared to the more volumetrically abundant and longer-lived ULX observed in the nearby universe.  This work is motivated not only by the potential of hyper-nebul\ae~as precursors to BH/NS common envelopes detectable by present or future radio surveys, but also by the previously hypothesized connection between these sources and the phenomena of extragalactic fast radio bursts (FRB; e.g., \citealt{Lorimer+07,Thornton+13}; \citealt{Petroff+19,Cordes&Chatterjee19} for reviews).  

Although most FRBs have only been detected once, several repeating FRB sources have now been discovered (e.g., \citealt{Spitler+16,CHIME_repeaters, James_19, Caleb+19, Ravi_19, Oostrum+20, Kirsten+22}). Two of the best-studied repeating sources, FRB 20180916 and \prsone, exhibit periodicities in the active windows over which bursts are detected of $\sim$16 and $\sim$160 days, respectively \citep{CHIME+20,Rajwade+20}.  Based in part on the similarities between these timescales and ULX super-orbital periods, \citet{Sridhar+21b} proposed ULX-like binaries as a source of repeating (periodic) FRBs.  In this scenario, FRBs are generated via magnetized shocks \citep{Lyubarsky14,Beloborodov17,Metzger+19} or magnetic reconnection \citep{Lyubarsky20,Mahlmann+22} in transient relativistic outflows\footnote{In essence, many of the physical processes that can take place within the magnetosphere or wind of magnetars (confirmed FRB sources; \citealt{CHIME_SGR1935,Bochenek_SGR1935paper}) can also occur in the magnetized relativistic jet of a BH or NS.} within the evacuated jet funnel of the accretion disk; periodicity in the burst activity window is imprinted by precession of the binary jet axis---along which the FRB is relativistically beamed---in and out of the observer's line of sight (see also \citealt{Katz17}). 

Two repeating FRB sources, \prsone~\citep{Chatterjee+17} and \prstwo~\citep{Niu+22}, are spatially co-located with luminous ($\nu L_{\nu} \gtrsim 3\times 10^{38}$\,erg\,s$^{-1}$) and compact $\lesssim 1$\,pc \citep{Marcote+17,Chen+22} sources of optically-thin synchrotron radio emission---so-called FRB persistent radio sources (PRS).  The bursts from both of these objects exhibit large and time-variable rotation measures, $|$RM$|\sim 10^{5}$\,rad\,m$^{-2}$, indicating that the source of the FRB emission is very likely embedded within the same dynamic, highly magnetized environment responsible for generating the synchrotron PRS (e.g., \citealt{Michilli+18,Plavin+22,Niu+22,Feng+22,Anna-Thomas+22,Mckinven+22}).  Previous works have attributed PRS variously to dwarf galaxy AGN (e.g., \citealt{Zhang17,Thompson19,Eftekhari+20,Wada+21}) or a plerion-like magnetized nebul\ae\ of relativistic particles powered by outflows from a young magnetar (e.g., \citealt{Kashiyama&Murase17,Metzger+17,Beloborodov17,Margalit&Metzger18,Zhao&Wang_21}; see also \citealt{Yang+20,Yang+22}).\footnote{Though we note that an electron-positron nebula, such as those inflated by the winds of rotation-powered pulsar or magnetar, cannot readily account for the high RM of known FRB PRS, which instead requires a magnetized medium with an electron-ion composition \citep{Michilli+18}.}  A related goal of the present work is therefore to assess whether PRS are consistent with being ULX hyper-nebul\ae, thereby supporting an accretion-jet origin for FRB emission and strengthening the potential connection between repeating FRB sources and future NS/BH common envelope events.

This paper is organized as follows. In Sec.~\ref{sec:model}, we outline a one-zone model for the radio nebulae of hyper-accreting sources.  We develop the model with analytical formalisms concentrating on the effect of slower outer disk winds in Sec.~\ref{subsec:disk_wind}, and the effect of the faster inner jet in Sec.~\ref{subsec:jet_wind}, on the dynamics of nebular expansion and radiation.  In Sec.~\ref{sec:results}, we present numerical solutions for the evolution of the particle distribution function in the nebula which predict its time-dependent observable properties, such as its radio light curve, spectra, and rotation measure.  In Sec.~\ref{sec:discussion}, we discuss the observational implications of our model: its detection prospects with blind surveys (Sec.~\ref{subsec:survey_detection}), application to FRBs (Sec.~\ref{subsec:frb_application}), and local-universe ULXs (Sec.~\ref{subsec:local_ULX}). We conclude in Sec.~\ref{sec:conclusion}, and provide a list of all the timescales introduced in this paper and their definitions in the Appendix.

\section{Model for ULX Hyper-Nebul\ae} \label{sec:model}

\begin{figure} 
\centering
\includegraphics[width=1.0\linewidth]{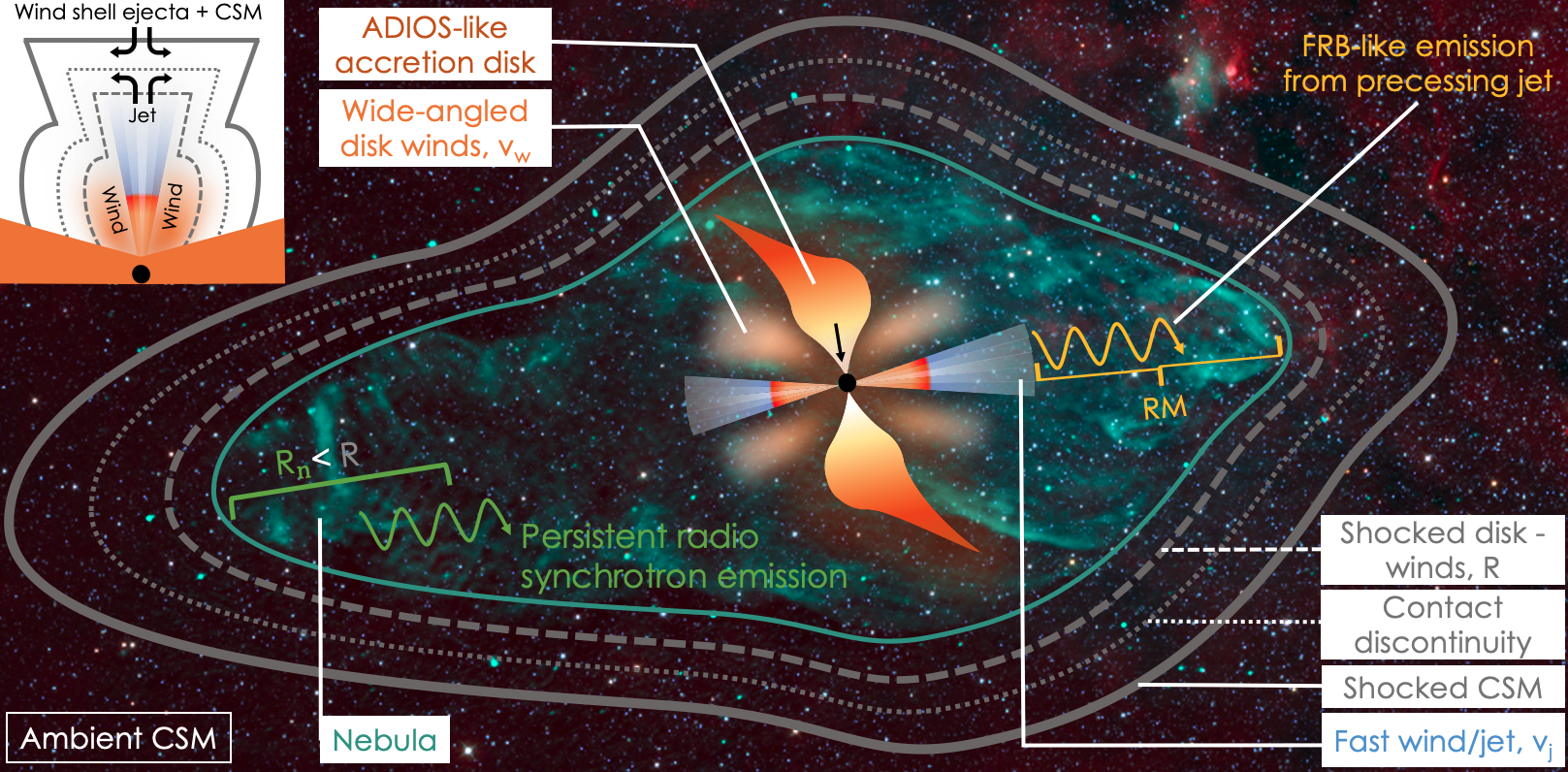}
\caption{Schematic diagram of the disk wind/jet-inflated nebula. The central black circle denotes a compact object (black hole or neutron star) accreting matter at near or exceeding super-Eddington rate from a companion star (not shown here) undergoing thermal- or dynamical-timescale mass transfer. The radiatively inefficient accretion disk `puffs-up' to resemble the advection dominated inflow-outflow solution (ADIOS; \citealt{Blandford&Begelman99}).  Such disks are subject to strong outflows in the form of wide-angled disk winds, which feed mass and energy into the large-scale environment, and helps shape the polar accretion funnel and the jet cavity.  The wind-fed ejecta shell drives a forward shock (solid grey curve) into the ambient circum-stellar medium (CSM, of density $\rho_{\rm csm}$) that is separated from the ejected shell/shocked disk-jet winds (dashed grey curve; at a distance $R$ given by Eq.~\ref{eq:R_fs}) by a contact discontinuity (dotted grey curve). The top-left inset shows how the jet `head' and the disk winds feed matter into their environment and shape-up their surroundings. Yellow wiggles emanating from the jet cavity represents an FRB-like coherent radio pulse beamed along the instantaneous jet axis (refer to \citealt{Sridhar+21b} for a detailed schematic and its emission mechanism).  The radio pulse travels through a nebula of electrons (green cloud) gyrating around the local magnetic field that imparts a large rotation measure (RM) to the pulse; the electrons cool via various radiative and expansion losses (Sec.~\ref{subsubsec:electron_evolution}), generating persistent synchrotron radio emission.  These electrons are energized and injected into the nebula by the shock formed at the jet - CSM/wind-shell interface (Sec.~\ref{subsec:jet_wind}). The location of this termination shock determines the boundary of the nebula (with a `size' $R_{\rm n}$; Eq.~\ref{eq:R_n}). The asymmetric/bipolar shape of the nebula is the result of the higher ram pressure of the jet/disk outflows along the polar accretion axis (with velocities $v_{\rm j}\gg v_{\rm w}$) and within the precession cone of the jet (possibly responsible for imparting periodicity in the active FRB phase).  The filamentary green structure underlying the schematic is an actual image of the Westerhout 50 (W50/`Manatee') nebula surrounding the Galactic ULX-like microquasar SS\,433, obtained with NRAO's Karl G. Jansky Very Large Array (VLA) and NASA’s Wide Field Survey Explorer (WISE; \citealt{WISE+10}). Credit for the actual image of the W50 nebula: NRAO/AUI/NSF, K. Golap, M. Goss.}
\label{fig:cartoon}
\end{figure}

In what follows, we outline a model for accretion-powered nebul\ae\ and their emission, motivated by observations of ULX bubbles which we briefly review here in order to justify the underlying physical picture (see Fig.~\ref{fig:cartoon} for a schematic illustration).

The X-ray luminosity of the high-inclination Galactic microquasar SS\,433 is only $\sim 10^{36-37}$\,erg\,s$^{-1}$ \citep{Abell&Margon_79}, but its face-on luminosity is predicted to be higher $\sim 10^{39-40}$\,erg\,s$^{-1}$, in which case it would likely resemble extragalactic ULX \citep{Margon84, Poutanen+07, Medvedev&Fabrika_10,Middleton+21}. 
Super-Eddington accretion flows of this type are geometrically thick, radiatively inefficient \citep{Abramowicz+80,Narayan&Yi95}, and dominated by mass outflows \citep{Blandford&Begelman99,Poutanen+07}. 
The latter include wide-angle disk winds and collimated jets, which inflate nebul\ae\ of thermal and non-thermal particles around the accretor, generating observable optical, X-ray, and radio signatures \citep{Abolmasov+09,Gurpide+22}.  Supporting this picture are Karl G. Jansky Very Large Array (VLA, henceforth) observations of the W50 `manatee' nebula surrounding the Galactic ULX-like microquasar SS 433, the nebula surrounding the ULX in NGC~7793 \citep{Dubner+98}, ATCA observations of the S26 nebula surrounding the ULX in NGC~7793 \citep{Soria+10}, and MUSE observations of NGC 1313 X–1 \citep{Gurpide+22}, for example.  

These and other observations suggest that ULX nebul\ae\ generally possess at least three distinct components: (i) large-scale quasi-spherical bubble inflated by the slow accretion disk winds \citep{Fabrika04,Pakull+06}, (ii) bipolar dumb-bell shaped lobes generated by the (often precessing) faster jet-wind activity \citep{Begelman+80,Velazquez&Raga_00,Zavala+08}, and (iii) an inner aspherical H\,{\sc ii} region due to the photoionizing X-rays radiated by the central accreting source \citep{Pakull&Mirioni02, Roberts+03, Abolmasov+09, Gurpide+22}. The slow winds from (i) interact with the circumstellar medium (CSM) via a forward shock, forming a shell of shocked CSM separated by the shocked wind material by a contact discontinuity \citep{Weaver+77}.  By contrast, the faster jet (source ii)---which is predicted to emerge along the instantaneous angular momentum axis of the inner accretion disk---interacts in a similar fashion via a termination shock with the swept up shell (source i); the latter exists even along the polar axis of the binary system due to the precession of the disk and hence the direction of its outflows.  The higher average velocity of the disk wind/jet material within the precession cone where they interact imparts ULX nebul\ae\ with their observed prolate (``egg shaped'') morphologies (e.g., \citealt{Pakull+10,Soria+10,Gurpide+22}). 

The bulk of the mechanical energy of the disk outflows is dissipated in forward and reverse shocks via source (i), which therefore governs the evolution of the overall size of the shell, and produces one source of electromagnetic emission \citep[thermal radio, optical, X-ray; e.g., ][]{Siwek+17}.  However, because the velocities of these shocks are relatively low, and their thermal and non-thermal radio luminosities relatively weak, considered alone they result in `radio-quiet' ULX bubbles.  By contrast, the shock interaction between the faster, relativistic jet and the slower wind-ejecta/CSM shell places a significantly power into relativistic electrons, generating radio synchrotron emission (seen as `lobes' or `hot spots' when resolved) which dominates over component (i) \citep[e.g., NGC 7793-S26;][]{Pannuti+02,Soria+10}.  In this paper we focus on such synchrotron `radio-loud' ULX bubbles, predominantly powered by the jet-shell interaction.

Some ULX also exhibit faster-varying non-thermal radio emission (e.g., \citealt{Middleton+13,Anderson+19b}), which originates from the base of the ULX jet.  However, for the extremely high $\dot{M}$-systems of greatest interest in this paper, the optical depth through such a high-luminosity jet (e.g. to synchrotron self absorption; SSA hereafter) may be too high for such emission from the jet base to escape.

\subsection{Binary Mass-Transfer Rate and Active Duration of Accretion Phase} 
\label{subsec:binary}

Consider a binary composed of a star of mass $M_{\star}$ transferring material via RLOF onto a BH or NS remnant of mass $M_{\bullet}$.  The source been accreting at close to the present rate for a time $t < t_{\rm active}$, where here $t_{\rm active}$ is the total duration of the accretion phase set by the mechanism of binary angular momentum loss and the response of the stellar envelope to mass-loss.   

If the donor star is undergoing stable mass-transfer on or close to the main-sequence, the duration of the accretion phase is roughly given by the donor's thermal timescale (e.g., \citealt{Kolb98,Sridhar+21b}),
\be \label{eq:t_active}
t_{\rm active} \sim t_{\rm th} \sim 10^{5}\,{\rm yr}\left(\frac{M_{\star}}{30M_{\odot}}\right)^{-1.6},\,\,\,\,\text{(Stable mass-transfer)}.
\ee
If the donor has evolved off the main sequence, the stable mass-transfer timescale can be even shorter, e.g. $t_{\rm active} \sim10^4$\,yr (or much longer in the case of nuclear-timescale mass-transfer; e.g., \citealt{Klencki+21}).\footnote{Short-timescale $t_{\rm active} \lesssim 10^{2}-10^{4}$ yr mass-transfer can also take place in the case of very massive donor stars that remain inflated after recently undergoing a pulsational pair-instability outburst (e.g., \citealt{Marchant+19}).}  On the other hand, the mass-transfer phase that immediately precedes a stellar merger or common envelope event is dynamically unstable, occurring in runaway fashion over a timescale as short as tens or hundreds of orbits (e.g., \citealt{Pejcha+17,MacLeod&Loeb20}):
\be \label{eq:t_active_orbperiod}
t_{\rm active} \sim (10-100)P \sim 3-30\,{\rm yr}\,\left(\frac{P}{100\,{\rm d}}\right), \,\,\,\,\text{(Unstable mass-transfer),}
\ee
where $P \sim 10-100$ days for typical ULX binary orbital periods (e.g., \citealt{Abell&Margon_79,Grise+13,Bachetti+14,Motch+14,Walton+16}).\footnote{Even shorter timescales $\lesssim$ days characterize accretion onto a stellar-mass black hole following the tidal disruption of a passing star (so-called ``micro-TDEs''; e.g., \citealt{Perets+16,Kremer+19}).  Although such events may generate high-energy transients, the high densities of the outflows from such sources render the ejecta opaque to the sources of emission explored in this paper.} 
For a given active time, the resultant binary mass-transfer rate is approximately,
\be
\dot{M} \approx \frac{M_{\star}}{t_{\rm active}}; \,\,\,\,\frac{\dot{M}}{\dot{M}_{\rm Edd}} \approx  10^{5}\left(\frac{M_{\star}}{30M_{\odot}}\right)\left(\frac{M_{\bullet}}{10M_{\odot}}\right)^{-1}\left(\frac{t_{\rm active}}{10^{3}{\rm \,yr}}\right)^{-1} \gg 1,
\label{eq:Mdot}
\ee
where, $\dot{M}_{\rm Edd} \equiv L_{\rm Edd}/(0.1\,c^{2})$ is the Eddington accretion rate and $L_{\rm Edd} \simeq 1.5\times 10^{39}(M_{\bullet}/10M_{\odot})$\,erg\,s$^{-1}$.    

Fig.~\ref{fig:timescales} summarizes $t_{\rm active}$ and other key timescales in the problem discussed below, as a function of the binary mass-transfer rate $\dot{M}$.  In order to explain observed FRB luminosities $L_{\rm FRB} \sim 10^{40}-10^{44}$\,erg\,s$^{-1}$ (top-left panel of Fig.~\ref{fig:timescales}) via an accretion-powered jet requires a minimum accretion rate onto the NS/BH.  As discussed in \citet{Sridhar+21b}, accounting for mass-loss in disk winds (which reduces the accreted mass reaching the inner disk by a factor $f_{\rm w} \sim 10^{-2}$ relative to the mass-transfer rate; \citealt{Blandford&Begelman99}) and the low predicted radio efficiencies of FRB emission models ($f_{\xi}\sim10^{-3}$; e.g., \citealt{Plotnikov&Sironi19,Sironi+21,Mahlmann+22}), this condition requires a binary mass-transfer rate,
\be \label{eq:L_FRB_Mdot}
\dot{M} \gtrsim 10^5 \dot{M}_{\rm Edd}\left(\frac{L_{\rm FRB}}{10^{42}\,{\rm erg\,s^{-1}}}\right)\left(\frac{f_{\rm w}}{10^{-2}}\cdot\frac{f_{\xi}}{10^{-3}}\right)^{-1}\left(\frac{f_{\rm b}}{10^{-2}}\right)\left(\frac{L_{\rm Edd}}{10^{39}\,{\rm erg}\,{\rm s}^{-1}}\right)^{-1}.
\ee
Here, $f_{\rm b}$ is the beaming factor, which is constrained in repeating FRB sources by the duty cycle of the burst activity (e.g., \citealt{Katz17,Sridhar+21b}).  Despite large uncertainties, Eqs.~\eqref{eq:Mdot}, \eqref{eq:L_FRB_Mdot} show that the observed FRB population is consistent with accreting systems characterized by a range of stable and unstable-mass transfer timescales, $t_{\rm active} \sim 10-10^{6}$\,yr.  

As an aside, we note that an upper limit on the mass-transfer rate arises from the requirement that the super-Eddington accretion disk must `fit' within the binary orbit (e.g., \citealt{King&Begelman99}).  For sufficiently high $\dot{M}$, mass-loss occurs through the outer $L_{2}$ Lagrangian point in an equatorially-concentrated circumbinary outflow \citep{Pejcha+16,Lu+22}, effectively limiting that feeding the NS/BH accretion flow.  The ``trapping radius'', interior to which the accretion rate is locally super-Eddington, is given by \citep{Begelman79}, 
\be
R_{\rm tr}\simeq \frac{2GM_{\bullet}}{c^2}\left(\frac{\dot{M}}{\dot{M}_{\rm Edd}}\right) \approx 4R_{\odot}\left(\frac{\dot{M}}{10^{5}\dot{M}_{\rm Edd}}\right). 
\ee
For main sequence stars, $\dot{M}$ is limited to relatively modest values $\lesssim 10^{5}\dot{M}_{\rm Edd}$ before $R_{\rm tr}$ exceeds the outer edge of the BH/NS accretion disk (typically comparable to the size of the binary orbit $\sim R_{\odot}$) and the fraction of the donor's mass-loss rate which feeds the outer disk (and hence the central NS/BH) drops due to $L_2$ mass-loss (e.g., \citealt{Lu+22}).  By contrast, for evolved giant star donors, larger values $\dot{M}/\dot{M}_{\rm Edd}\gtrsim 10^7$ are possible because of the larger orbital separation of the binary $\gtrsim 10^{2} R_{\odot}$. 

\subsection{Disk Wind-Inflated Nebula} \label{subsec:disk_wind}

The disk wind outflows that accompany highly super-Eddington accretion carry a total mass loss-rate $\dot{M}_{\rm w}$ which nearly equals the entire mass-transfer rate, i.e.~$\dot{M}_{\rm w} \sim \dot{M}$ (e.g., \citealt{Blandford&Begelman99,Hashizume+15}), with only a small fraction making its way down to the BH/NS surface.  Depending on the radial scale of their launching point in the disk or binary, the outflows can possess a range of speeds, from values similar to the binary orbital velocity, $v_{\rm orb} \lesssim 100$ km s$^{-1}$, to the trans-relativistic speeds $v_{\rm j} \gtrsim 0.3\,c$ which characterize jet-like outflows from the innermost radii of the disk (and similar to those observed in SS\,433 and ULX; e.g., \citealt{Jeffrey+16} and references therein).  In advection dominated inflow-outflow solution (ADIOS) models for the radial disk structure \citep{Blandford&Begelman99,Margalit&Metzger16} the mass accretion rate decreases as a power-law with radius $\dot{M} \propto r^{p}$ for $r \in [r_{\rm in},r_{\rm out} \lesssim R_{\rm tr}]$, where $r_{\rm out} \gg r_{\rm in}$.  Assuming the outflows local to each annulus in the disk reach an asymptotic velocity equal to the local disk escape velocity, the kinetic energy-averaged wind velocity is given by (e.g., \citealt{Metzger12})
\be
v_{\rm w} \approx \sqrt{\frac{p}{1-p}\frac{GM_{\bullet}}{r_{\rm in}}\left(\frac{r_{\rm in}}{r_{\rm out}}\right)^{p}} \approx 0.05\,{\rm c}\,\left(\frac{M_{\bullet}}{10M_{\odot}}\right)^{0.3}\left(\frac{r_{\rm in}}{6r_{\rm g}}\right)^{-0.2}\left(\frac{r_{\rm out}}{R_{\odot}}\right)^{-0.3},
\label{eq:vw}
\ee
where in the second equality we have normalized $r_{\rm in}$ to gravitational radii $r_{\rm g} = GM_{\bullet}/c^{2}$ and take $p = 0.6$ (as supported by hydrodynamical simulations of radiatively-inefficient accretion flows; e.g., \citealt{Yuan&Narayan14,Hu+22}).  In what follows, we take $v_{\rm w} \sim 0.03\,c$ as a fudicial value, consistent with that expected for a main sequence or moderately evolved donor star feeding a relatively compact accretion flow, $r_{\rm out} \sim R_{\odot}$.  However, lowers value $v_{\rm w} \lesssim 0.01\,c$, may be appropriate for a giant donor star ($r_{\rm out} \gtrsim 10^{2}R_{\odot}$) or for larger values of $p$.  The kinetic luminosity of the wind is then 
\be
L_{\rm w} \approx \frac{1}{2}\dot{M}_{\rm w}v_{\rm w}^{2}  \approx 10^{42}\,{\rm erg\,s^{-1}}\left(\frac{\dot{M}_{\rm w}}{10^{5}\dot{M}_{\rm Edd}}\right)\left(\frac{M_{\bullet}}{10M_{\odot}}\right)v_{\rm w,9}^{2}\,{\rm erg\,s^{-1}},
\label{eq:Lw}
\ee
where $v_{\rm w,9} = v_{\rm w}/(10^{9}$ cm s$^{-1}$) and hereafter we adopt the short-hand notation $Y_{\rm x} \equiv Y/10^{\rm x}$ for quantities given in cgs units. In analytic estimates hereafter, we shall typically fix the value of $M_{\bullet} = 10M_{\odot}$ and use $L_{\rm w}$ and $\dot{M}_{\rm w}$ interchangeably. 

The quasi-spherical disk winds expand into the circumstellar medium (CSM) of assumed constant density $\rho_{\rm csm} = \mu n m_{\rm p}$, where $\mu=1.38$ is the mean atomic weight for neutral solar composition gas, and we adopt a density $n\approx 10$ cm$^{-3}$ typical of the environments of massive stars \citep{Abolmasov+08,Pakull+10,Toala&Arthur_2011}.  Initially, the wind expands freely into the CSM, until sweeping up a mass comparable to its own, as occurs at the radius $r_{\rm cs}.$  Equating the mass released in the wind $\approx \dot{M}_{\rm w}t$ up to a given time $t \sim r_{\rm cs}/v_{\rm w}$ to the mass of the swept-up CSM, $\approx (4\pi/3)r_{\rm cs}^{3}\rho_{\rm csm}$, \be
r_{\rm cs} = \left(\frac{L_{\rm w}}{2\pi \rho_{\rm csm}v_{\rm w}^{3}}\right)^{1/2} \approx 0.7\,{\rm pc}\,\left(\frac{L_{\rm w,42}}{n_{1}v_{\rm w,9}^3}\right)^{1/2}.
\ee 
The free-expansion phase thus lasts a duration,
\be
t_{\rm free} \approx \frac{r_{\rm cs}}{v_{\rm w}} \approx 70\,{\rm yr}\,\left(\frac{L_{\rm w,42}}{n_{1}}\right)^{1/2}v_{\rm w,9}^{-5/2}.
\label{eq:tfree}
\ee
For stable mass-transfer systems, $t_{\rm active} \gtrsim 10^{5}$ yr $\gg t_{\rm free}$, i.e. the system is active well past the free expansion phase.  By contrast, for the unstable case $t_{\rm free} \gtrsim t_{\rm active} \sim 1-100$ yr, i.e. the wind ejecta released preceding a common envelope event is likely to be freely expanding.  

At later times $t \gg t_{\rm free}$, the ejecta shell begins to appreciably decelerate as its swept-up mass exceeds that injected by the wind, causing the shell radius to grow more gradually in time, $R \propto t^{3/5}$ \citep{Weaver+77}.  The radius evolution, bridging the free expansion and decelerating phases, can thus be approximated as
\begin{equation}\label{eq:R_fs}
    R(t) \simeq
    \begin{cases}
          v_{\rm w}t \approx 0.7\,{\rm pc} \left(\frac{t}{70\,{\rm yr}}\right) & (t < t_{\rm free}) \\
          \alpha\left(\frac{L_{\rm w}t^{3}}{\rho_{\rm csm}}\right)^{1/5} \approx 0.8\,{\rm pc} \,\left(\frac{L_{\rm w,42}}{n_{1}}\right)^{1/5}\left(\frac{t}{70\,{\rm yr}}\right)^{3/5} & (t > t_{\rm free}),\\
    \end{cases}
\end{equation}
where $\alpha \approx 0.88$ while the forward shock is adiabatic and $\alpha \approx 0.76$ after it becomes radiative (\citealt{Weaver+77}; as occurs on a timescale $\sim 10^{4}-10^{5}$ yr, see Sec.~\ref{subsubsec:X-ray}).  After the central outflow source turns off at $t > t_{\rm active}$, the nebula's expansion transitions to a more rapid Sedov-Taylor deceleration phase, for which $R \propto t^{3/5}$.

The mass accumulated inside the shell is 
\begin{equation} \label{eq:Msh}
 M_{\rm sh} \sim 
    \begin{cases}
          \dot{M}_{\rm w}t \approx 1.5M_{\odot} \left(\frac{\dot{M}_{\rm w}}{10^5 \dot{M}_{\rm Edd}}\right)\left(\frac{t}{70\,{\rm yr}}\right)& (t < t_{\rm free}) \\
          (4\pi/3)\rho_{\rm csm}R^{3} \approx 0.8M_{\odot} \left(L_{\rm w,42}^3n_1^2t_{9.3}^9\right)^{1/5} & (t > t_{\rm free}).
    \end{cases}
\end{equation}
Assuming the shell remains ionized, its expansion results in a (maximum) dispersion measure (DM) through the shell,
\begin{equation}\label{eq:DM}
    \text{DM}_{\rm sh} \simeq \frac{M_{\rm sh}}{4\pi R^{2}m_{\rm p}} \approx
    \begin{cases}
          10\,{\rm pc\,cm^{-3}}\left(\frac{\dot{M}_{\rm w}}{10^{5}\dot{M}_{\rm Edd}}\right)v_{\rm w,9}^{-2}\left(\frac{t}{70\,{\rm yr}}\right)^{-1} & (t < t_{\rm free}) \\
          4\,{\rm pc\,cm^{-3}}\left(\frac{L_{\rm w,42}}{n_{1}}\right)^{1/5}\left(\frac{t}{70\,{\rm yr}}\right)^{3/5} & (t > t_{\rm free}).\\
    \end{cases}
\end{equation}

\begin{figure} 
\begin{minipage}{.5\textwidth}
  \begin{subfigure}
    \centering
    \includegraphics[width=1.20\linewidth]{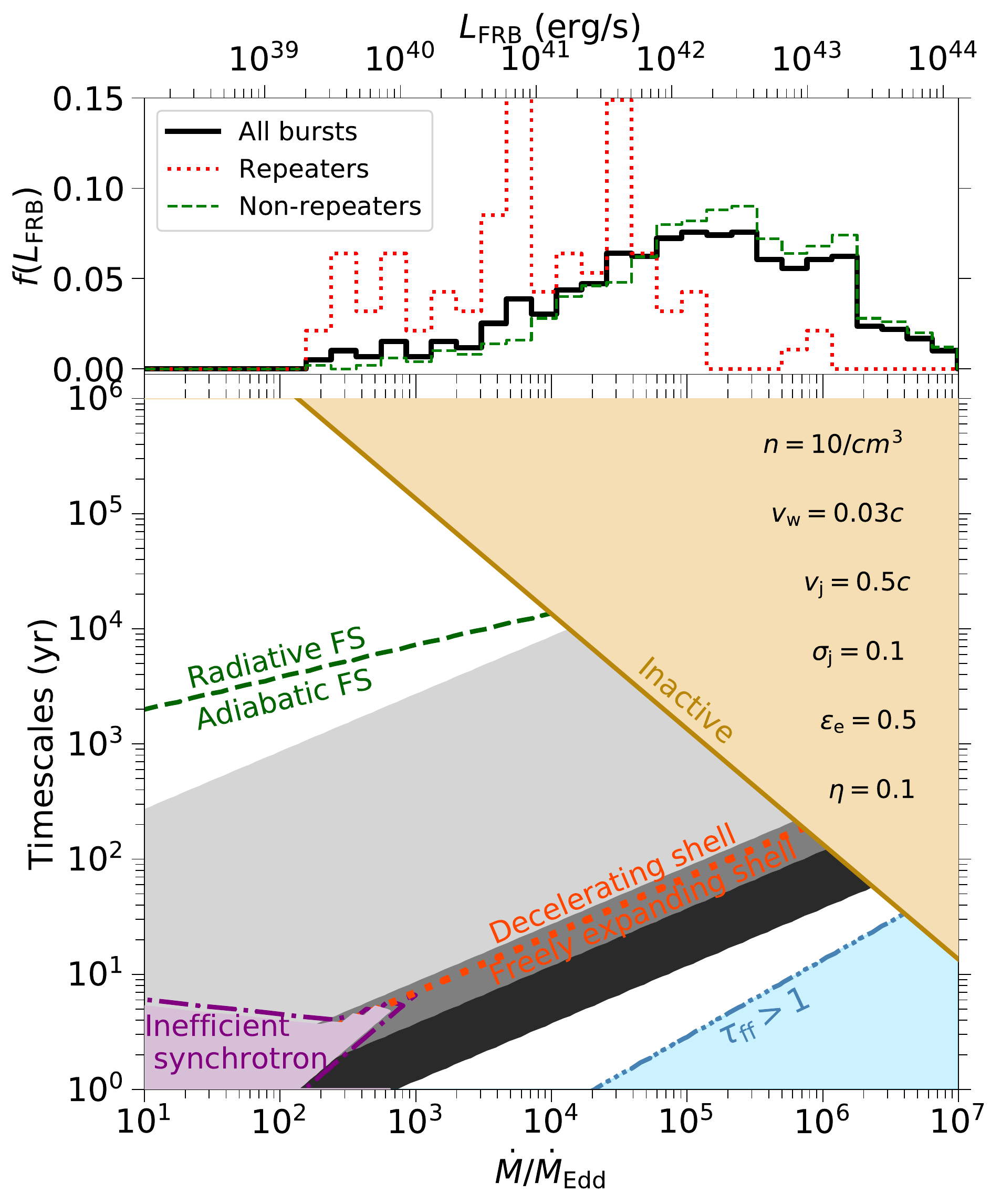}
    \label{fig:sub3}
  \end{subfigure}
\end{minipage}%
\hspace{1.5cm}
\begin{minipage}{.5\textwidth}
  \begin{subfigure}
    \centering
    \includegraphics[width=.82\linewidth]{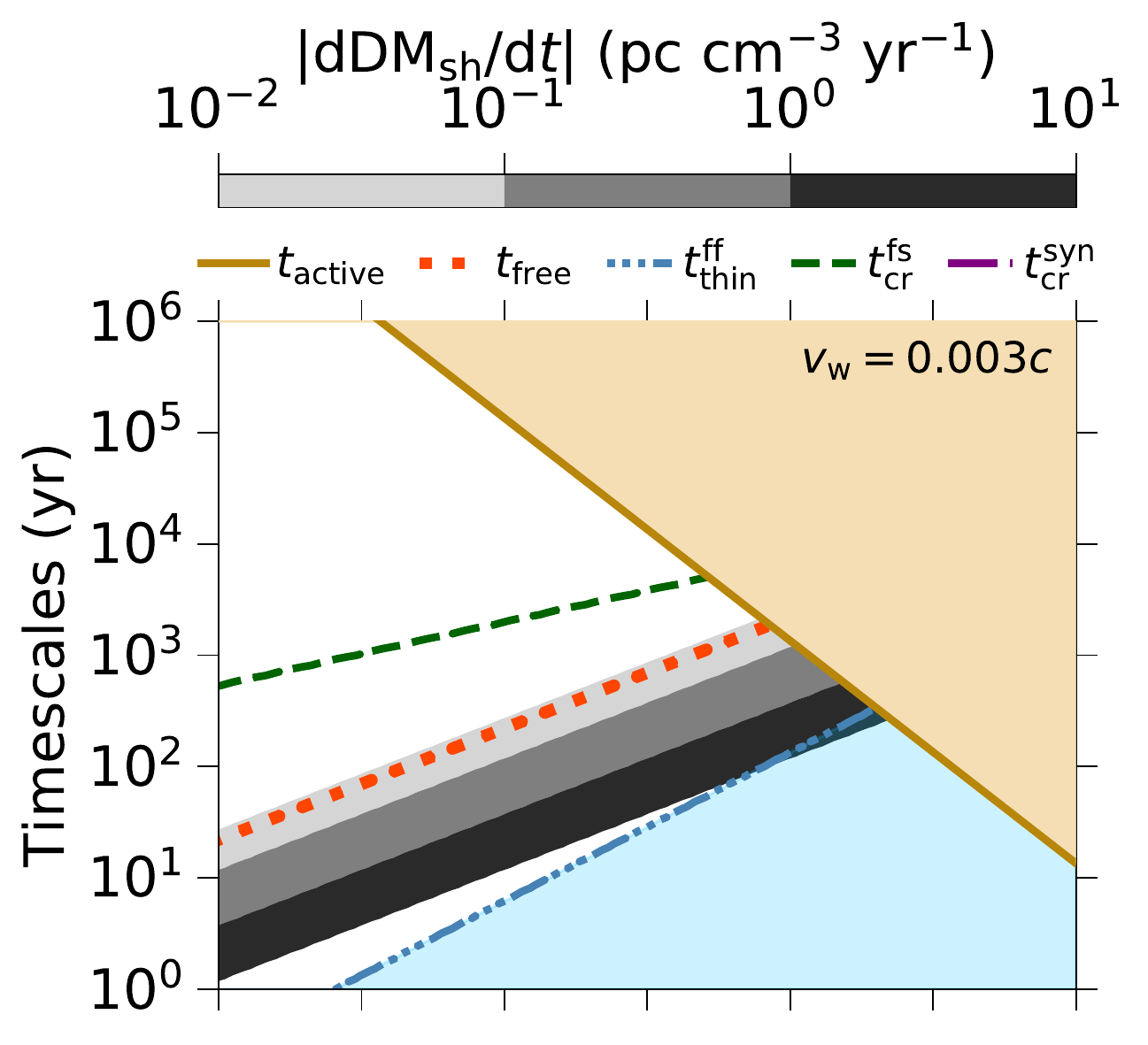}
    \label{fig:sub1}
  \end{subfigure}\\
  \begin{subfigure}
    \centering
    \includegraphics[width=.81\linewidth]{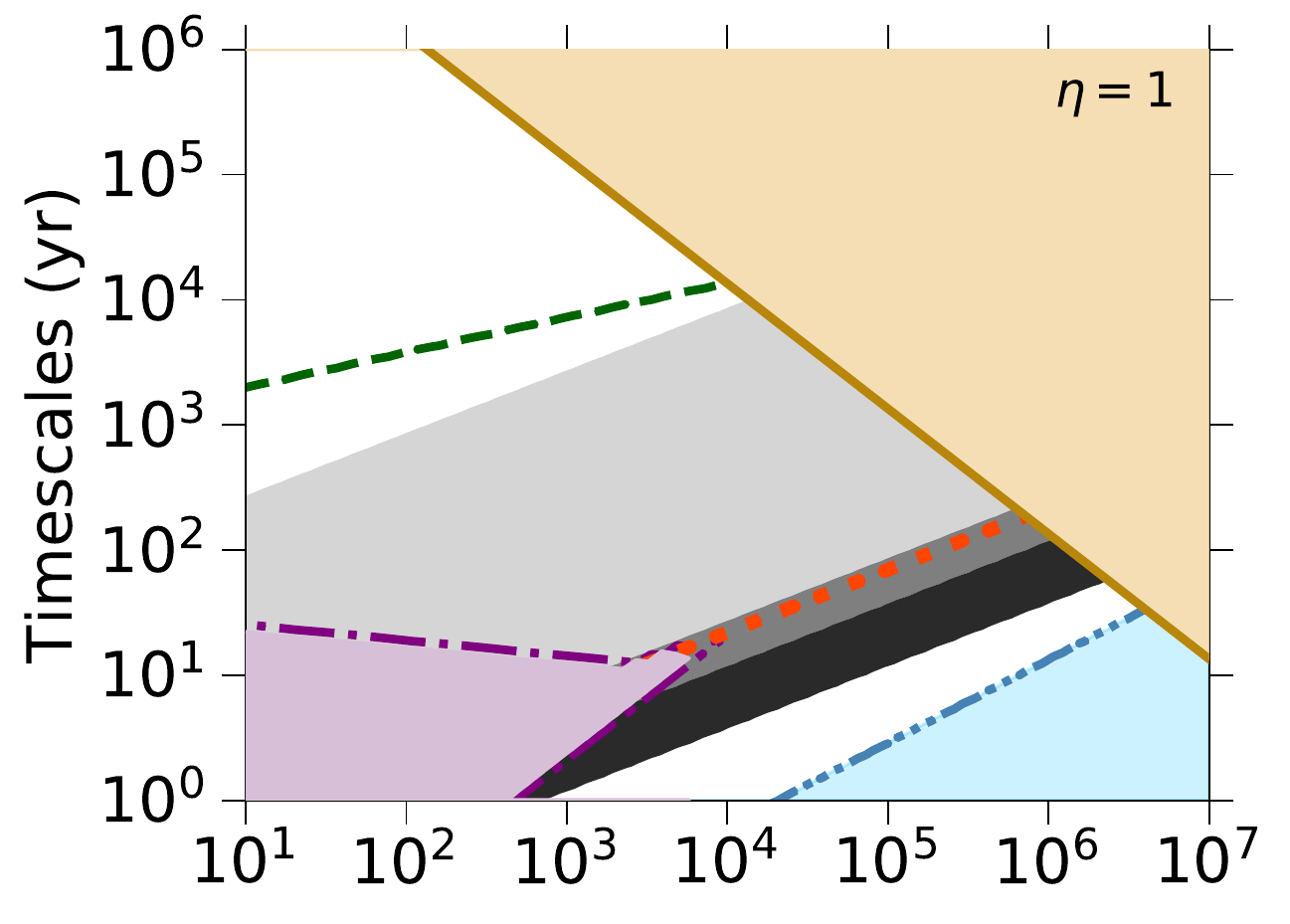}
    \label{fig:sub2}
  \end{subfigure}
\end{minipage}%
\caption{Critical timescales in the evolution of ULX hyper-nebul\ae\ as a function of the mass-transfer rate $\dot{M}$ in units of the Eddington rate $\dot{M}_{\rm Edd} \equiv L_{\rm Edd}/(0.1 c^{2})$ for an assumed accretor mass $M_{\bullet} = 10M_{\odot}$ and donor mass $M_{\star} = 30M_{\odot}$. The bottom-left panel is computed for the fiducial model (parameters listed along the top-right side), and the right panels present the critical timescales for models with a different $v_{\rm w}=0.003\,c$ (top) and $\eta=1$ (bottom), but otherwise the same set of remaining fiducial parameters. Different lines and regions denote the following. (1) brown-solid line: maximum active time $t_{\rm active}$ (Eq.~\ref{eq:Mdot}); (2) red-dotted line: duration of free-expansion phase, $t_{\rm free}$ (Eq.~\ref{eq:tfree}); (3) blue-dot-dot-dotted line: the time $t>t_{\rm thin}^{\rm ff}$ after which radio emission (at 1\,GHz) is no longer attenuated by free-free absorption (i.e., $\tau_{\rm ff}<1$; Eq.~\ref{eq:t_ffthin}); (4) green-dashed line: the time $t>t_{\rm cr}^{\rm fs}$ after which the forward shock becomes radiative (Eq.~\ref{eq:ffcool_exp_timescale}); (5) purple dash-dotted line: thermal electrons entering nebula cool efficiently via synchrotron emission at times $t<t_{\rm cr}^{\rm syn}$ during free expansion regime and $t>t_{\rm cr}^{\rm syn}$ during decelerating regime (Eq.~\ref{eq:syncool_exp_timescale}). Grey shaded regions show the maximum time derivative of the dispersion measure ($\text{DM}_{\rm sh}$) through the wind ejecta shell (Eq.~\ref{eq:DM}). Note that synchrotron cooling is efficient for the entire presented range of duration (1-$10^6$\,yr) and $\dot{M}/\dot{M}_{\rm Edd}$ ($10^1-10^7$) for the $v_{\rm w}=0.003\,c$ model. The top-left panel is the fractional FRB luminosity distribution calculated from the first CHIME catalog; solid black histogram represents all the bursts, red-dotted histogram represents the first burst from repeaters, and the green-dashed histogram represents the non-repeaters \citep{CHIME_catalog_21}: the FRB luminosity is along the top horizontal axis and the minimum required mass-transfer rate along the bottom axis (following Eq.~\ref{eq:L_FRB_Mdot}, adopting the fiducial but uncertain values for the radiative efficiency and beaming).}
\label{fig:timescales}
\end{figure}

Gray contours in Fig.~\ref{fig:timescales} show $|{\rm d} \text{DM}_{\rm sh}/{\rm d}t|$ as a function of $\dot{M}$ and source age.  Other potential sources of time-dependent DM include electrons in the nebula (Sec.~\ref{subsubsec:electron_evolution}) and the unshocked disk outflow at radii $\ll R$, driven by the orbital motion or precession of the inner disk (\citealt{Sridhar+21b}).  The optical depth through the shell to free-free absorption is approximately given by,
\begin{equation}\label{eq:tau_ff}
    \tau_{\rm ff} \approx \alpha_{\rm ff}R(t) \approx
    \begin{cases}
          7\times10^{-5}\nu_{9}^{-2}T_{4}^{-3/2}v_{\rm w,9}^{-5}\left(\frac{\dot{M}_{\rm w}}{10^5~\dot{M}_{\rm Edd}}\right)^{2}\left(\frac{t}{70\,{\rm yr}}\right)^{-3} & (t < t_{\rm free}) \\
          9\times10^{-6}\nu_{9}^{-2}T_{4}^{-3/2}v_{\rm w,9}^{-2}\left(\frac{\dot{M}_{\rm w}}{10^5~\dot{M}_{\rm Edd}}\right)\left(\frac{t}{70\,{\rm yr}}\right)^{-1}  & (t > t_{\rm free}),\\
    \end{cases}
\end{equation}
where $\alpha_{\rm ff} \approx 0.018Z^{2}\nu^{-2}T^{-3/2}n_{\rm e,sh} n_{\rm ion,sh}\bar{g}_{\rm ff}$ cm$^{-1}$ \citep{Rybicki&Lightman_79}, $\nu$ is the observing frequency, $\bar{g}_{\rm ff}\sim1$ is the Gaunt factor, and we take $Z \sim 1$, $n_{\rm e,sh} \simeq n_{\rm i,sh} \simeq 3M_{\rm sh}/(4\pi R^{3} m_{\rm p})$ for the electron and ion density in the shell, respectively (i.e., we have assumed a shell thickness $\sim R$, a reasonable approximation during the early phases when the forward shock is adiabatic and the absorption is most relevant).  This estimate of $\tau_{\rm ff}$ is an upper limit because we have again assumed the shell to be fully ionized along the line of sight, e.g. by the central X-ray source, with a temperature $T \sim 10^{4}$ K.  The ejecta becomes optically thin to free-free emission ($\tau_{\rm ff} < 1$) after a time,
\be \label{eq:t_ffthin}
t_{\rm thin}^{\rm ff} \approx 2.8\,{\rm yr}~ \nu_{9}^{-2/3}v_{\rm w,9}^{-5/3}T_{4}^{-1/2}\left(\frac{\dot{M}_{\rm w}}{10^5~\dot{M}_{\rm Edd}}\right)^{2/3},
\ee
where we have self-consistently assumed $t < t_{\rm free}.$  Fig.~\ref{fig:timescales} shows that for $\sim$GHz observing frequencies, $t_{\rm thin}^{\rm ff}$ is the shortest timescale in the system evolution.  Any FRB emission from the vicinity of the central binary, or synchrotron emission from the nebula embedded behind the shell, will thus be free to escape for most of the source's active lifetime.

\subsubsection{Thermal X-rays from Ejecta/CSM Shock Interaction} \label{subsubsec:X-ray}

\begin{figure} 
\centering
\includegraphics[width=0.7\linewidth]{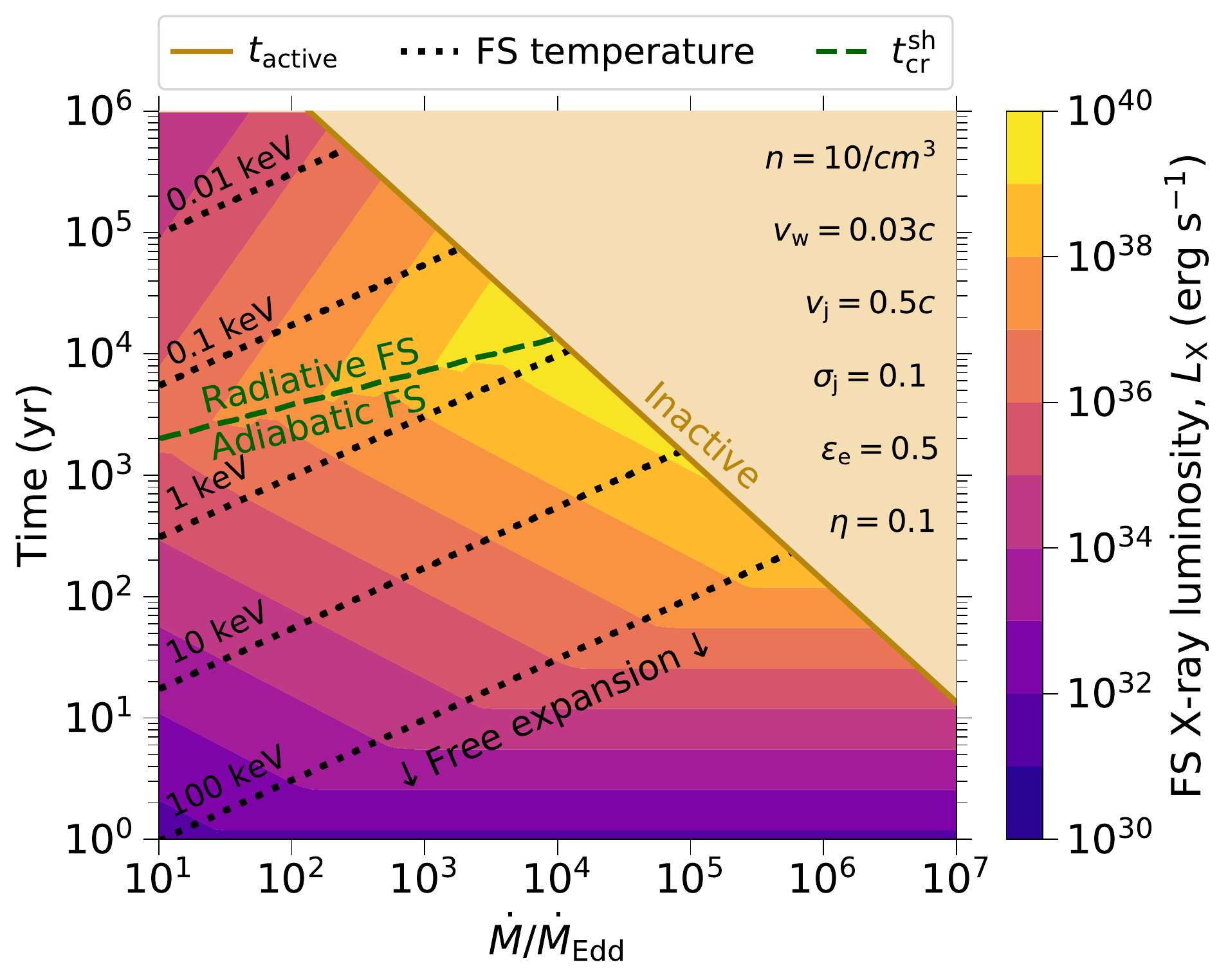}
\caption{Colored contours show the thermal X-ray luminosity (Eq.~\ref{eq:L_X}) powered by the forward shock of ULX nebul\ae\ as a function of the mass transfer rate $\dot{M}/\dot{M}_{\rm Edd}$ for the same parameters as in the left panel of Fig.~\ref{fig:timescales}.  The forward shock initially expands freely, before beginning to decelerate at $t > t_{\rm free}$ (Eq.~\ref{eq:tfree}).  Black dotted contours denote the temperature of the gas heated behind the forward shock (Eq.~\ref{eq:X-ray_temp}).  
}
\label{fig:L_X}
\end{figure}

Most of the thermal X-ray emission from ULX nebul\ae\ (e.g., \citealt{Pakull+10}) are produced by the disk wind shell-CSM shock interaction (e.g., \citealt{Siwek+17}).  We focus on emission from the forward shock (FS), which becomes radiative first and hence will typically dominate over the luminosity of the reverse shock (RS), except at hard X-ray energies.  

The FS heats gas to a temperature set by the shock jump conditions,
\be \label{eq:X-ray_temp} 
kT_{\rm fs}=\frac{3}{16}\mu m_{\rm p} v_{\rm fs}^2\approx130\,{\rm keV}\left(\frac{L_{\rm w,42}}{n_{1}}\right)^{2/5}\left(\frac{t}{70\,{\rm yr}}\right)^{-4/5},
\ee
where we take $v_{\rm fs} = {\rm d}R/{\rm d}t = 3R/5t$ for the FS velocity, corresponding to the deceleration phase $t > t_{\rm free}$ (Eq.~\ref{eq:R_fs}).  The cooling time of the shocked gas is (e.g., \citealt{Vlasov+16}),
\be \label{eq:t_fs_cool}
t_{\rm cool}^{\rm fs} \simeq \frac{3kT_{\rm fs}}{2n_{\rm fs}\mu\Lambda[T_{\rm fs}]}\approx 1\times10^{6}\,{\rm yr}\left(\frac{L_{\rm w,42}}{n_{1}}\right)^{2/5}\left(\frac{t}{70\,{\rm yr}}\right)^{-2/5},
\ee
where $n_{\rm fs}=4n$ is the post-shock density for adiabatic index $\hat{\gamma}=5/3$, and we have assumed a cooling function appropriate for solar-metallicity gas that includes both free-free emission and lines $\Lambda[T] = \Lambda_{\rm ff} + \Lambda_{\rm line}$, where \citep{Schure+09,Draine11},
\begin{subequations} \label{eq:cooling_function}
\begin{gather}
\Lambda_{\rm ff}[T] \simeq 3\times10^{-27}(T/{\rm K})^{0.5}\,{\rm erg~s}^{-1}{\rm cm}^{3},\\
\Lambda_{\rm lines}[T] \simeq 2.8\times 10^{-18}(T/{\rm K})^{-0.7}\,{\rm erg~s}^{-1}{\rm cm}^{3}.
\end{gather}
\end{subequations}

The ratio of the particle cooling time to the expansion timescale $t_{\rm exp} \sim (5/3)t$, is thus given by
\be \label{eq:ffcool_exp_timescale}
\frac{t_{\rm cool}^{\rm fs}}{t_{\rm exp}} \approx 1\times10^4\left(\frac{L_{\rm w,42}}{n_{1}}\right)^{2/5}\left(\frac{t}{70\,{\rm yr}}\right)^{-7/5},
\ee
where we have assumed free-free cooling dominates ($\Lambda \sim \Lambda_{\rm ff}$), as is valid for $T_{\rm fs} \gtrsim 10^{7.3}$ K.  The FS thus becomes radiative ($t_{\rm cool}^{\rm fs} < t_{\rm exp}$) after a critical time 
\be \label{eq:t_fs_cr}
t_{\rm cr}^{\rm fs} \approx 5\times10^4\,{\rm yr}\left(\frac{L_{\rm w,42}}{n_{1}}\right)^{2/7},
\ee
as shown with a green-dashed line in Fig.~\ref{fig:timescales}.  

After the radiative transition $t \gtrsim t_{\rm cr}^{\rm fs}$, the total luminosity $L_{\rm fs}$ radiated behind the FS will follow its kinetic luminosity,
\be 
L_{\rm fs} \approx L_{\rm kin,fs} \approx \frac{9\pi}{8} \rho_{\rm fs} R^2 v_{\rm fs}^3\approx 6\pi R^{2}v_{\rm fs}n kT_{\rm fs}\approx7\times10^{41}\,{\rm erg\,s}^{-1}\,L_{\rm w,42}, ~~~~~~ (t > t_{\rm cr}^{\rm fs}),
\ee
where $\rho_{\rm fs}[R] = \rho_{\rm csm}$. Before the radiative transition $t < t_{\rm cr}^{\rm fs}$, the radiated luminosity is suppressed from this maximal value, roughly according to (e.g., \citealt{Vlasov+16}),
\be
\label{eq:L_fs}
L_{\rm fs} \approx \left(1+\frac{5}{2} \frac{t_{\rm cool}^{\rm fs}}{t_{\rm exp}}\right)^{-1}L_{\rm kin,fs} \underset{t_{\rm exp} > t_{\rm cool}^{\rm fs}}\approx 7\times10^{41} {\rm erg~s}^{-1} \left(1+\frac{5}{2} \frac{t_{\rm cool}^{\rm fs}}{t_{\rm exp}}\right)^{-1} L_{\rm w,42}, ~~~~~~ (t < t_{\rm cr}^{\rm fs}).
\ee
Free-free emission from the shock will peak at photon energies $h\nu_{\rm X} \sim kT_{\rm fs}$, typically in the X-ray band.  We therefore estimate the X-ray luminosity as the portion of the shock's total luminosity emitted via free-free emission:
\be \label{eq:L_X}
L_{\rm X,fs} \approx \left(\frac{\Lambda_{\rm ff}}{\Lambda}\right)L_{\rm fs}.
\ee
This is a conservative lower limit on $L_{\rm X}$ because it assumes all line emission is radiated in the optical/UV instead of X-rays. 

In addition to the FS, the RS becomes strong at times $t > t_{\rm free}$ and will produce its own free-free emission of temperature
\be
kT_{\rm rs} \simeq \frac{3}{16}\mu m_{\rm p} (v_{\rm w}-v_{\rm fs})^{2} \approx \frac{3}{16}\mu m_{\rm p} v_{\rm w}^{2} \approx 200\,{\rm keV}\,v_{\rm w,9}^{2}
\ee
and luminosity
\be \label{eq:L_rs}
L_{\rm rs}  \approx \left(1+\frac{5}{2} \frac{t_{\rm cool}^{\rm rs}}{t_{\rm exp}}\right)^{-1}L_{\rm kin,rs} \underset{t_{\rm exp} > t_{\rm cool}^{\rm rs}}\approx 4\times10^{41} {\rm erg~s}^{-1} \left(1+\frac{5}{2} \frac{t_{\rm cool}^{\rm fs}}{t_{\rm exp}}\right)^{-1} L_{\rm w,42}, ~~~~~~ (t < t_{\rm cr}^{\rm fs}),
\ee
where the kinetic luminosity of the RS is given by,
\be
L_{\rm kin,rs}\approx \frac{9\pi}{8}\rho_{\rm w}[R]R^{2}v_{\rm w}^{3} = \frac{9}{32}\dot{M}v_{\rm w}^{2}\approx 4\times10^{41}\,{\rm erg~s}^{-1} L_{\rm w,42},
\ee
$\rho_{\rm w}[R] = \dot{M}/(4\pi v_{\rm w}R^{2})$ and
\be \label{eq:t_rs_cool}
t_{\rm cool}^{\rm rs} \simeq \frac{3kT_{\rm rs}}{2n_{\rm rs}\mu\Lambda_{\rm ff}[T_{\rm rs}]}\approx 2\times10^6\,{\rm yr}\left(\frac{L_{\rm w,42}}{n_{1}}\right)^{2/5}\left(\frac{t}{70\,{\rm yr}}\right)^{-2/5},
\ee
where $n_{\rm rs} = 4\rho_{\rm w}/(\mu m_{\rm p})$ is the density of the shocked wind.  The luminosities of the forward and reverse shocks (Eqs.~\ref{eq:L_fs}, \ref{eq:L_rs}) are comparable at $t\sim t_{\rm free}$ because the temperatures and densities of the shocked gas that determine their cooling timescales are similar at this epoch (e.g., $\rho_{\rm csm}\sim\rho_{\rm w}$) at $t\sim t_{\rm free}$.  However, at later times $t\gg t_{\rm free}$ (and especially after the FS becomes radiative) we have $t_{\rm cool}^{\rm rs}\gg t_{\rm cool}^{\rm fs}$, resulting in $L_{\rm fs} \gg L_{\rm rs}$.  For these reasons, we hereafter neglect the RS and focus attention on the FS emission.

Fig.~\ref{fig:L_X} shows the FS X-ray luminosity for the same parameters and ranges of $\dot{M}$ and system age as in Fig.~\ref{fig:timescales}. The highest X-ray luminosity $L_{\rm X}\sim 10^{40}$\,erg\,s$^{-1}$ is attained for systems accreting at $\dot{M}/\dot{M}_{\rm Edd}\sim 10^4$, after the FS becomes radiative at time $t \sim t_{\rm cr}^{\rm fs} \sim 10^{4}$ years.  For $\dot{M}/\dot{M}_{\rm Edd} \ll 10^3$ the peak luminosity is lower and attained at later times ($t \sim t_{\rm cr}^{\rm fs}$), while for systems with much higher $\dot{M}/\dot{M}_{\rm Edd} \gg 10^4$, the system becomes inactive before the FS becomes radiative.  The peak emission temperature drops from $kT \sim$100\,keV at $t \sim t_{\rm free}$ to $kT \sim 0.1$\,keV by the time the X-ray luminosity peaks at $t \sim t_{\rm cr}^{\rm fs}$. Due to this, at $t > t_{\rm cr}^{\rm fs}$, the forward shock is dominated by line cooling, and the free-free X-ray luminosity starts decreasing. As we shall discuss in Sec.~\ref{sec:discussion}, these X-ray luminosities are likely to be challenging to detect at the typically large distances of ULX hyper-nebul\ae. 

\subsection{Jet-Inflated Radio Nebula} \label{subsec:jet_wind}

Slower winds from the disk ($v_{\rm w} \sim 0.03\,c$; Eq.~\ref{eq:vw}) dominate the total mass-loss from the binary, but outflows from the innermost regions of the disk reach much higher velocities ($v_{\rm j} \gtrsim 0.3\,c$) and hence dominate the radio synchrotron emission (e.g., \citealt{Urquhart+18}).  We hereafter refer to this mildly relativistic outflow region as the ``jet'', even though (1) it may be distinct from the cleaner ultra-relativistic outflow (e.g.~one originating in the neutron star magnetosphere or threading the black hole horizon) capable of generating FRB emission \citep{Sridhar+21b}; (2) it is an idealization to divide the disk outflows cleanly into distinct ``slow'' and ``fast'' components; a gradual radial gradient in the outflow properties is more physically realistic.  

We take the luminosity of the jet,
\be
L_{\rm j} = \frac{1}{2}\dot{M}_{\rm j} v_{\rm j}^{2} = \eta L_{\rm w}, 
\ee
to be a fraction $\eta$ of the total disk wind power $L_{\rm w}$ (Eq.~\ref{eq:Lw}), where the mass-loss rate of the jet therefore obeys $\dot{M}_{\rm j}= \eta\dot{M}_{\rm w}(v_{\rm w}/v_{\rm j})^2$.  Values of $\eta \gtrsim 0.1$ are predicted in ADIOS models, due to the roughly equal gravitational energy released per radial decade in the accretion flow (e.g., \citealt{Blandford&Begelman99}).  

The collision between the jet and the CSM/disk-wind shell described in the previous section, inflates a bipolar nebula of relativistically hot particles and magnetic fields behind the shell.  Equating the jet ram-pressure $\sim \dot{M}_{\rm j} v_{\rm j}/4\pi R_{\rm n}^{2} = L_{\rm j}/(2\pi R_{\rm n}^{2}v_{\rm j})$ with the thermal pressure of the nebula $\sim 3L_{\rm w}t_{\rm exp}/4\pi R^{3} \sim 3L_{\rm w}/(4\pi R^{2}v)$, we estimate the characteristic nebula radius
\be R_{\rm n} \sim R\left(\frac{2vL_{\rm j}}{3v_{\rm j}L_{\rm w}}\right)^{1/2} \approx
\begin{cases}
    0.05\,{\rm pc}\,v_{\rm w,9}^{3/2} \left(\frac{v_{\rm j}}{0.5\,c}\right)^{-1/2} \eta^{1/2}_{-1} \left(\frac{t}{70\,{\rm yr}}\right) & (t<t_{\rm free}; v=v_{\rm w})\\
    0.05\,{\rm pc}\, \left(\frac{L_{\rm w,42}}{n_1}\right)^{3/10} \left(\frac{v_{\rm j}}{0.5\,c}\right)^{-1/2} \eta^{1/2}_{-1} \left(\frac{t}{70\,{\rm yr}}\right)^{2/5}& (t>t_{\rm free}; v=v_{\rm fs}).
\end{cases} \label{eq:R_n}
\ee
Although the nebula shape is bipolar rather than spherical, we take its total volume to be $V_{\rm n} = 4\pi R_{\rm n}^3/3$; the larger polar radius of the shell compared to its average will partially compensate for the limited latitudinal extent, e.g. as set by the misalignment angle of the jet precession cone (Fig.~\ref{fig:cartoon}).

A termination shock separates the unshocked jet from the nebula, where the electrons are heated to relativistic energies.  In what follows, we present a one-zone model for the nebular electrons from which we calculate their synchrotron radio emission and estimate the RM and DM through the nebula.  This model is adapted from that of \citet{Margalit&Metzger18} who applied it in the different context of nebul\ae\ inflated by the ejecta of a flaring magnetar.

\subsubsection{Electron Evolution and Synchrotron Radiation} \label{subsubsec:electron_evolution}

The number density of electrons in the nebula $N_\gamma {\rm d}\gamma$ with Lorentz factors between $\gamma$ and $\gamma+{\rm d}\gamma$, evolves in time according to the continuity equation
\be \label{eq:continuity_eq}
\dot{N}_\gamma = \frac{\partial}{\partial t}N_\gamma + \frac{\partial}{\partial\gamma}(\dot{\gamma}N_\gamma) - 3\frac{\dot{R}_{\rm n}}{R_{\rm n}}N_\gamma,
\ee
where the second term on the right hand side account for losses for adiabatic expansion and radiation, to be enumerated below.

The source term $\dot{N}_\gamma$ in Eq.~\eqref{eq:continuity_eq} accounts for the injection of fresh electrons into the nebula at the jet termination shock.  Assuming an electron-ion composition of the jet, electrons heated at the shock will enter the nebula with a mean Lorentz factor (e.g., \citealt{Margalit&Metzger18})
\be
\bar{\gamma}_{\rm e} \simeq \frac{1}{2}\varepsilon_{\rm e}\frac{m_{\rm p}}{m_{\rm e}}\frac{v_{\rm j}^{2}}{c^{2}} \approx 115\left(\frac{\varepsilon_{\rm e}}{0.5}\right)\left(\frac{v_{\rm j}}{0.5\,c}\right)^{2},
\label{eq:gammath}
\ee
where $\varepsilon_{\rm e}$ is the heating efficiency of the electrons (values of $\varepsilon_{\rm e} \sim 0.1-0.5$ are found in particle-in-cell simulations of magnetized shocks spanning non-relativistic to transrelativistic speeds; \citealt{Sironi&Spitkovsky11,Tran&Sironi20}).  The energy distribution of the thermal electrons is assumed to be a relativistic Maxwellian of temperature $kT = \bar{\gamma}_{\rm e} m_{\rm e} c^{2}/3$, with a total particle injection rate obeying
\be \label{eq:injection_rate}
\dot{N}_{\rm e,th} = \frac{4\pi R_{\rm n}^3}{3}\int \dot{N}_\gamma d\gamma = \frac{L_{\rm j}}{2\bar{\gamma_{\rm e}}m_{\rm e}c^2(1+\sigma_{\rm j})},
\ee  
where $\sigma_{\rm j} \lesssim 1$ is the jet magnetization (the ratio of its Poynting flux to kinetic energy flux).  

Although we neglect such a possibility in this paper, an additional non-thermal power-law distribution of electrons could be added to the injected population at this stage in the calculation.  This may be necessary to model the radio emission from (lower-$\dot{M}$) ULX in the nearby universe, for which the thermal synchrotron peak is below the typical observing frequencies and power-law synchrotron spectra are measured for the jet hot-spots (e.g., \citealt{Urquhart+18}; Sec.~\ref{subsec:local_ULX}). 

If only as a result of the jet material being magnetized, the nebula will be magnetized, with an average magnetic field strength $B_{\rm n}$.  The magnetic energy $E_{\rm B} = (B_{\rm n}^{2}/8\pi)V_{\rm n}$ of the nebula is assumed to evolve in time according to
\be
\frac{{\rm d}E_{\rm B}}{{\rm d}t} = \left(\frac{\sigma_{\rm j}}{1+\sigma_{\rm j}}\right)L_{\rm j} - \frac{\dot{R}_{\rm n}}{R_{\rm n}}E_{\rm B},
\label{eq:dEBdt}
\ee
where the first term in the right hand side accounts for the injection of magnetic fields from the jet and the second term accounts for adiabatic losses (assuming the magnetic field is tangled and evolves as a gas of an effective adiabatic index $\hat{\gamma} = 4/3$).  Equation (\ref{eq:dEBdt}) assumes that the magnetic energy is not dissipated in the nebula faster than the expansion timescale, which is tantamount to the assumption of a constant nebula magnetization, $\sigma_{\rm n}= (v_{\rm j}/c)^2\sigma_{\rm j} \sim 10^{-2}(\sigma_{\rm j}/0.1)$, where $\sigma_{\rm n}=B_{\rm n}^2R_{\rm n}^3/(3\dot{M}_{\rm j}c^2t)$.  A similar magnetization $\sigma_{\rm n} \sim 10^{-2}$ is inferred for the Crab Nebula from its synchrotron emission and axial ratio (e.g., \citealt{Kennel&Coroniti84,Begelman&Li92}).  

Energy losses of electrons with Lorentz factor $\gamma = (1-\beta^{2})^{-1/2}$ and velocity $\beta \equiv v/c$ are captured in Eq.~\eqref{eq:continuity_eq} via the loss term,
\be \label{eq:gdot}
\dot{\gamma}=\dot{\gamma}_{\rm ad} + \dot{\gamma}_{\rm brem} + \dot{\gamma}_{\rm IC} + \dot{\gamma}_{\rm syn}, 
\ee
which includes contributions from adiabatic expansion \citep{Vurm&Metzger_18},
\be
\dot{\gamma}_{\rm ad} = -\frac{1}{3}\gamma\beta^2\frac{{\rm d}\ln{V_{\rm n}}}{{\rm d}t} = -\gamma\beta^2\frac{\dot{R}_{\rm n}}{R_{\rm n}},
\ee
bremsstrahlung emission,
\be
\dot{\gamma}_{\rm brem} = -\frac{5}{3}c\sigma_{\rm T}\alpha_{\rm fs}n_{\rm e}\gamma^{1.2},
\ee
where $n_{\rm e} = \int N_{\gamma}d\gamma$ is the electron density, $\alpha_{\rm fs}\simeq1/137$ is the fine-structure constant and, synchrotron and inverse-Compton radiation,
\be \label{eq:sync_IC_loss}
\dot{\gamma}_{\rm syn,IC} = -\frac{4}{3}\frac{\sigma_{\rm T}}{m_{\rm e}c}\beta^2\gamma^2
    \begin{cases}
         f_{\rm ssa}B_{\rm n}^2/8\pi & ({\rm synchrotron}) \\
         L_{\rm tot}/4\pi cR_{\rm n}^2 & ({\rm inverse{-}Compton}),\\
    \end{cases}
\ee
where $L_{\rm tot} = \int L_{\nu} d\nu.$  Given the distribution of electrons $N_{\gamma}$ and nebula magnetic field $B_{\rm n}$, the synchrotron luminosity is calculated according to
\be \label{eq:synch_luminosity}
L_{\nu} = 4\pi^2R_{\rm n}^2\frac{j_\nu}{\alpha_\nu}(1-e^{-\alpha_\nu R_{\rm n}}),
\ee
\be \label{eq:emission_absorption}
j_\nu = \int\frac{N_\gamma P_\nu(\gamma)}{4\pi}d\gamma,~~~\alpha_\nu = -\int\frac{\gamma^2P_{\nu}(\gamma)}{8\pi m_{\rm e}\nu^2}\frac{\partial}{\partial\gamma}\left[\frac{N_\gamma}{\gamma^2}\right]d\gamma,
\ee
where 
\be\label{eq:spectral_power}
P_{\nu} = \frac{2e^3B_{\rm n}}{\sqrt{3}m_{\rm e}c^2}F\left(\frac{\nu}{\nu_{\rm syn}}\right),~~~F(x) \equiv x\int_{x}^{\infty}K_{5/3}(y)dy,
\ee
is the spectral power of a synchrotron-emitting electron \citep{Rybicki&Lightman_79}, where $K_{\rm 5/3}(y)$ is the modified Bessel function of order 5/3, and 
\begin{eqnarray}
\nu_{\rm syn} = \frac{eB_{\rm n}}{2\pi m_{\rm e} c}\gamma^{2}
\label{eq:nusynth}
\end{eqnarray}
is the characteristic synchrotron frequency.  We follow \citet{Margalit&Metzger18} in modifying the synchrotron cooling loss term (Eq.~\ref{eq:sync_IC_loss}) to account for SSA by multiplying it by a suppression factor,
\be \label{eq:f_ssa}
f_{\rm ssa}(\gamma) \approx \frac{1-e^{-\tau(\gamma)}}{\tau(\gamma)} < 1,
\ee
where $\tau(\gamma) = \alpha_\nu\nu_{\rm syn,th}R_{\rm n}$ is the typical optical depth through the nebula of the synchrotron emission from an electron of energy $\gamma$, where $\nu_{\rm syn,th}$ is the characteristic emission frequency of the electrons (Eq.~\ref{eq:nusynth} evaluated for $\gamma = \bar{\gamma}_{\rm e}$; Eq.~\ref{eq:gammath}).  Finally, to account for free-free absorption through the wind shell at early times, we attenuate the intrinsic radio luminosity according to,
\be \label{eq:ff_absorption}
L_{\nu} \rightarrow L_{\nu}e^{-\tau_{\rm ff}(\nu)},
\ee
where $\tau_{\rm ff}(\nu)$ is given by Eq.~\eqref{eq:tau_ff}.

In addition to tracking the time-dependent DM through the ejecta shell (Eq.~\ref{eq:DM}) and nebula,
\be
\text{DM}_{\rm neb} \simeq R_{\rm n} \int\frac{N_{\gamma}}{\gamma}d\gamma   ,
\ee
for each model we calculate the magnitude of the RM through the nebula according to
\be
|\text{RM}| \simeq \frac{e^{3}}{2\pi m_{\rm e}^{2}c^{4}}\left(\frac{\lambda}{R_{\rm n}}\right)^{1/2}B_{\rm n}R_{\rm n}\int \frac{N_{\gamma}}{\gamma^{2}}d\gamma,
\label{eq:RM}
\ee
where $\lambda \le R_{\rm n}$ is the effective coherence length of the magnetic field inside the nebula (e.g., \citealt{Margalit&Metzger18}; see further discussion in the next section).  

\subsubsection{Analytic Estimates of Synchrotron Emission and Rotation Measure} \label{subsubsec:analytic_sync_RM}

Before proceeding to the numerical results, here we provide analytic estimates for the nebular synchrotron emission and RM.  
From Eq.~\eqref{eq:dEBdt}, the nebula magnetic field strength can be estimated from the injected magnetic energy $E_{\rm B} \sim \sigma_{\rm j} L_{\rm j}t_{\rm exp}$ over the expansion timescale $t_{\rm exp} \sim t$,
\begin{equation}\label{eq:B_n}
    B_{\rm n} \simeq \left(\frac{6\sigma_{\rm j} L_{\rm j}t}{R_{\rm n}^{3}}\right)^{1/2} \approx
    \begin{cases}
          0.18\,{\rm G}\,\sigma_{\rm j,-1}^{1/2}\eta_{-1}^{-1/4}\left(\frac{\dot{M}_{\rm w}}{10^5\,\dot{M}_{\rm Edd}}\right)^{1/2}v_{\rm w,9}^{-5/4}\left(\frac{v_{\rm j}}{0.5\,c}\right)^{3/4}\left(\frac{t}{70\,{\rm yr}}\right)^{-1} & (t < t_{\rm free}) \\
          0.18\,{\rm G}\,\sigma_{\rm j,-1}^{1/2}\eta_{-1}^{-1/4}\left(\frac{\dot{M}_{\rm w}}{10^5\,\dot{M}_{\rm Edd}}\right)^{1/20}v_{\rm w,9}^{1/10}\left(\frac{v_{\rm j}}{0.5\,c}\right)^{3/4}n_{1}^{9/20}\left(\frac{t}{70\,{\rm yr}}\right)^{-1/10} & (t > t_{\rm free}).\\
    \end{cases}
\end{equation}
The synchrotron peak frequency of the electrons of energy $\bar{\gamma}_{\rm e}$ (Eq.~\ref{eq:gammath} for $\varepsilon_{\rm e} = 0.5$) is then, 
\begin{eqnarray}
\nu_{\rm syn,th} &\approx& 0.3\nu_{\rm syn}(\bar{\gamma}_{\rm e}) \approx
\begin{cases}
1.9\,{\rm GHz} \,\sigma_{\rm j,-1}^{1/2}\eta_{-1}^{-1/4}\left(\frac{\dot{M}_{\rm w}}{10^5\,\dot{M}_{\rm Edd}}\right)^{1/2}v_{\rm w,9}^{-5/4}\left(\frac{v_{\rm j}}{0.5\,c}\right)^{17/4}\left(\frac{t}{70\,{\rm yr}}\right)^{-1} & (t < t_{\rm free})\\
2.0\,{\rm GHz} \,\sigma_{\rm j,-1}^{1/2}\eta_{-1}^{-1/4}\left(\frac{\dot{M}_{\rm w}}{10^5\,\dot{M}_{\rm Edd}}\right)^{1/20}v_{\rm w,9}^{1/10}\left(\frac{v_{\rm j}}{0.5\,c}\right)^{17/4}n_{1}^{9/20}\left(\frac{t}{70\,{\rm yr}}\right)^{-1/10} & (t > t_{\rm free}).
\label{eq:nusynthnum}
\end{cases}
\end{eqnarray}
The radiative cooling time of the thermal electrons is, 
\begin{eqnarray} \label{eq:t_cool_syn}
t_{\rm cool}^{\rm syn} \simeq \frac{6\pi m_{\rm e} c}{\sigma_{\rm T}B_{\rm n}^{2}\bar{\gamma}_{\rm e}} & \approx & \frac{10(m_{\rm e} e c)^{1/2}}{\sigma_{\rm T}B_{\rm n}^{3/2}\nu_{\rm syn,th}^{1/2}} \approx
    \begin{cases}
        7.3\,{\rm yr}\,\sigma_{\rm j,-1}^{-1}\eta_{1/2}^{-1}\left(\frac{\dot{M}_{\rm w}}{10^5\,\dot{M}_{\rm Edd}}\right)^{-1}v_{\rm w,9}^{5/2}\left(\frac{v_{\rm j}}{0.5\,c}\right)^{-5/2}\left(\frac{t}{70\,{\rm yr}}\right)^{2} & (t < t_{\rm free}) \\
        6.5\,{\rm yr}\,\sigma_{\rm j,-1}^{-1}\eta_{1/2}^{-1}\left(\frac{\dot{M}_{\rm w}}{10^5\,\dot{M}_{\rm Edd}}\right)^{-1/10}v_{\rm w,9}^{-1/5}\left(\frac{v_{\rm j}}{0.5\,c}\right)^{-5/2}n_1^{-9/10}\left(\frac{t}{70\,{\rm yr}}\right)^{1/5} & (t < t_{\rm free})
    \end{cases}       
\end{eqnarray}
The ratio of the synchrotron cooling time to the nebula expansion timescale is given by,
\begin{eqnarray} \label{eq:syncool_exp_timescale}
\frac{t_{\rm cool}^{\rm syn}}{t_{\rm exp}} \approx
    \begin{cases}
        0.1\,\sigma_{\rm j,-1}^{-1}\eta_{1/2}^{-1}\left(\frac{\dot{M}_{\rm w}}{10^5\,\dot{M}_{\rm Edd}}\right)^{-1}v_{\rm w,9}^{5/2}\left(\frac{v_{\rm j}}{0.5\,c}\right)^{-5/2}\left(\frac{t}{70\,{\rm yr}}\right) & (t < t_{\rm free}) \\
        0.06\,\sigma_{\rm j,-1}^{-1}\eta_{1/2}^{-1}\left(\frac{\dot{M}_{\rm w}}{10^5\,\dot{M}_{\rm Edd}}\right)^{-1/10}v_{\rm w,9}^{-1/5}\left(\frac{v_{\rm j}}{0.5\,c}\right)^{-5/2}n_1^{-9/10}\left(\frac{t}{70\,{\rm yr}}\right)^{-4/5} & (t < t_{\rm free}).
    \end{cases}       
\end{eqnarray}
If $t_{\rm cool}^{\rm syn} \ll t_{\rm exp}$ the injected electrons have time to radiate most of their luminosity $L_{\rm e} = \varepsilon_{\rm e}L_{\rm j}$ at frequency $\sim \nu_{\rm syn,th}$ before they cool adiabatically.  This efficiency condition is obeyed at early and late times in the nebula evolution, 
\begin{eqnarray} \label{eq:t_syn_cr}
t
    \begin{cases}
        \ll t_{\rm cr}^{\rm syn} \simeq 700\,{\rm yr}\,\sigma_{\rm j,-1}\eta_{-1}\left(\frac{\dot{M}_{\rm w}}{10^5\,\dot{M}_{\rm Edd}}\right)v_{\rm w,9}^{-5/2}\left(\frac{v_{\rm j}}{0.5\,c}\right)^{5/2} & (t < t_{\rm free})\\
        \gg t_{\rm cr}^{\rm syn} \simeq 2\,{\rm yr}\,\sigma_{\rm j,-1}^{-5/4}\eta_{-1}^{-5/4}\left(\frac{\dot{M}_{\rm w}}{10^5\,\dot{M}_{\rm Edd}}\right)^{-1/8}v_{\rm w,9}^{-1/4}\left(\frac{v_{\rm j}}{0.5\,c}\right)^{-25/8}n_{1}^{-9/8} & (t > t_{\rm free}).
    \end{cases}       
\end{eqnarray}
As shown in Fig.~\ref{fig:timescales}, for fiducial assumptions (e.g., regarding $\varepsilon_{\rm e}$, $\sigma_{\rm j}$, $v_{\rm j}$), the fast-cooling phase includes most of the mass-transfer rates and system ages of interest.  

Finally, the RM through the nebula (Eq.~\ref{eq:RM}) can be approximated,
\begin{eqnarray} \label{eq:RM_analytical}
|\text{RM}| &\approx& f_{\rm cool}\frac{e^{3}}{2\pi m_{\rm e}^{2}c^{4}} \frac{3\dot{M}_{\rm j}t}{4\pi R_{\rm n}^{3}m_{\rm p}}B_{\rm n}R_{\rm n}\left(\frac{\lambda}{R_{\rm n}}\right)^{1/2}\nonumber \\
&\approx& 
    \begin{cases}
        4.1\times10^4\,{\rm rad\,m^{-2}}\, f_{\rm cool}\sigma_{\rm j,-1}^{1/2}\eta_{-1}^{1/2}\left(\frac{\dot{M}_{\rm w}}{10^5\,\dot{M}_{\rm Edd}}\right)^{3/2}v_{\rm w,9}^{-3/2}\left(\frac{v_{\rm j}}{0.5\,c}\right)^{-1}\left(\frac{t}{70\,{\rm yr}}\right)^{-2} & (t < t_{\rm free}) \\
        4.7\times10^4\,{\rm rad\,m^{-2}}\, f_{\rm cool}\sigma_{\rm j,-1}^{1/2} \eta^{1/20} \left(\frac{\dot{M}_{\rm w}}{10^5\,\dot{M}_{\rm Edd}}\right)^{21/20}v_{\rm w,9}^{21/20}\left(\frac{v_{\rm j}}{0.5\,c}\right)^{-1}n_{1}^{9/10}\left(\frac{t}{70\,{\rm yr}}\right)^{7/10} & (t > t_{\rm free}), 
    \end{cases}       
\end{eqnarray}
where $f_{\rm cool} < 1$ is the fraction of the injected electrons that have cooled through radiation or adiabatic expansion to sub-relativistic energies ($\gamma \beta \lesssim 1$) and hence appreciably contribute to the RM integral.  We have also used Eq.~\eqref{eq:Lw} to write $L_{\rm j} = \eta L_{\rm w} = (\eta/2)\dot{M}_{\rm w}v_{\rm w}^{2}$.

For a given nebular magnetic field and particle distribution, the maximum rotation measure, $|$RM$|_{\rm max}$, is obtained in the limit that the magnetic field is coherent across the entire nebula ($\lambda \sim R_{\rm n}$), while more generally we expect $|$RM$|<|$RM$|_{\rm max}$ if $\lambda < R_{\rm n}$.  On top of any secular decline or rise in the average value $|$RM$|$ for an average fixed value $\lambda/R_{\rm n}$ predicted by Eq.~\eqref{eq:RM_analytical}, we note that shorter-timescale fluctuations (including possible sign-flips) in RM are possible if the magnetic field structure inside the nebula is evolving (effectively, leading to a time-dependent $\lambda/R_{\rm n}$).  In analogy with other magnetized nebul\ae\ fed by the termination shock of a relativistic wind/jet, such as pulsar wind nebul\ae, changes in the nebular magnetic field structure could arise due to vorticity (e.g., \citealt{Porth+13}) or turbulence  (e.g., \citealt{Zrake&Arons_17,Bucciantini+17}) generated near the termination shock.  The characteristic timescale for such $|$RM$|$ fluctuations could be as short as the shock's dynamical timescale,
\be 
t_{\rm \delta RM} \sim \frac{R_{\rm n}}{v_{\rm j}} \simeq \begin{cases}
    0.32\,{\rm yr}\,v_{\rm w,9}^{3/2} \left(\frac{v_{\rm j}}{0.5\,c}\right)^{-3/2} \eta^{1/2}_{-1} \left(\frac{t}{70\,{\rm yr}}\right) & (t<t_{\rm free})\\
    0.32\,{\rm yr}\, \left(\frac{\dot{M}_{\rm w}}{10^5\,\dot{M}_{\rm Edd}}\right)^{3/10} v_{\rm w,9}^{3/5} \left(\frac{v_{\rm j}}{0.5\,c}\right)^{-3/2} \eta^{1/2}_{-1} n_1^{-3/10} \left(\frac{t}{70\,{\rm yr}}\right)^{2/5}& (t>t_{\rm free}),
\end{cases} \label{eq:t_deltaRM}
\ee
i.e. $\sim$weeks to months in the youngest accreting sources of age $t \sim 10-100$ years.  Changes in the RM could in principle also be driven by variations of the line of sight from the FRB source through the nebula, e.g. as driven by precession of the inner accretion funnel along which the FRB emission is beamed \citep{Sridhar+21b}.  However, any such periodic RM-dependence will likely be washed out in favor of more stochastic variations unless $t_{\rm \delta RM}$ is long compared to the precession period.

\subsection{Summary of the model} \label{subsec:summary_model}

In summary, the main parameter of the model is the active timescale $t_{\rm active}$ (equivalently, mass-transfer rate $\dot{M}$), which can vary by orders of magnitude between binary systems depending on the mass-transfer mechanism (e.g., thermal or dynamical timescale; Sec.~\ref{subsec:binary}).  The system evolution is also sensitive to other uncertain parameters: $\eta \equiv L_{\rm j}/L_{\rm w} \in [0.1,1], \sigma_{\rm j} \in [0.01,1], v_{\rm w}/c \in [0.003-0.03], v_{\rm j}/c \in [0.1-1], \varepsilon_{\rm e} \in [0.2,0.8]$ whose values we shall vary, as well as several auxiliary parameters whose values we typically fix: $ M_{\star} = 30M_{\odot}, M_{\bullet} = 10 M_{\odot}, n = 10\,{\rm cm^{-3}}$.  The model predicts the observed radio luminosity $L_{\nu}$, DM$_{\rm sh}$, DM$_{\rm neb}$ and $|$RM$|_{\rm max} = |$RM$|(\lambda/R_{\rm n} = 1)$ as a function of time $t$ since the onset of accretion activity.  

As shown by Fig.~\ref{fig:timescales}, a hierarchy of timescales is generally expected, with $t_{\rm thin}^{\rm ff} < t_{\rm free} < t_{\rm cr}^{\rm syn} < t_{\rm cr}^{\rm fs}$, though only some of these phases are achieved within the active time of the source, depending on $\dot{M}$.

\section{Numerical Results} \label{sec:results}

The formalism developed in the previous sections is numerically solved here over a range of system's parameters.  Each of these models provides the temporal evolution of the observable properties of the expanding ejecta/nebula. The physical parameters of the hyper-nebula ($R_{\rm n}$, $\dot{R_{\rm n}}, N_\gamma$, $B_{\rm n}$) are first independently evolved assuming a free-expansion, and a self-similar expansion profile, separately across all times. These two profiles are then numerically stitched i.e., the resulting modest numerical discontinuity at $t_{\rm free}$, if any, is then bridged by employing a smoothing function to obtain a single trajectory of physical parameter smoothly evolving in time. For example,
\begin{subequations} \label{eq:cooling_function}
\begin{gather}
R_{\rm n} = R_{\rm n, free} \left[1-f_{\rm smooth}(t, t_{\rm free})\right] + R_{\rm n, ss} f_{\rm smooth}(t, t_{\rm free}), \\
f_{\rm smooth}(t, t_{\rm free}) = \frac{1}{2}\left\{1 + \tanh{\left[0.05(t-t_{\rm free})\right]}\right\},
\end{gather}
\end{subequations}
where the nebular radius $R_{\rm n, free}$ assumes a free-expansion profile, and $R_{\rm n, ss}$ assumes a self-similar expansion profile. Individual nebular parameters are integrated and evolved in time using `upwind differencing' scheme, which then yield the loss-terms and observables in Sec.~\ref{subsubsec:electron_evolution}. We turn off the central engine at $t=t_{\rm active}$ by multiplying the wind luminosity by $f_{\rm smooth}(t,t_{\rm active})$.

\subsection{Example Models for Fixed $\dot{M}$}
\label{sec:singleMdot}

We begin in Sec.~\ref{subsec:nebular_properties} by describing in detail a fiducial model, corresponding to a relatively high mass-transfer rate ($\dot{M}/\dot{M}_{\rm Edd}=10^5$) with a fixed active lifetime of $t_{\rm active}\approx1.3\times10^3$\,yr. The other parameters of our fiducial model are, $v_{\rm w}=0.03\,c$, $v_{\rm j}=0.5\,c$, $\sigma_{\rm j}=0.1$, $\eta=0.1$, and $\varepsilon_{\rm e}=0.5$ (see also Sec.~\ref{subsec:summary_model}).  Then, Sec.~\ref{subsubsec:observable_properties} covers the time-evolving observable properties of the nebula (viz., light curve, spectral energy distribution, $|{\rm RM}|$, and $\text{DM}_{\rm neb}$).  In addition to the fiducial model, here we examine the changes in the observable properties that arise by varying $v_{\rm w}/c \in [0.003,0.03]$, $v_{\rm j}/c \in [0.1,1]$, $\sigma_{\rm j} \in [0.01,1],$ and $\eta \in [0.1,1]$, about their fiducial values.

\subsubsection{Intrinsic Nebular Properties} \label{subsec:nebular_properties}

\begin{figure} 
\begin{center}
    \includegraphics[width=1\linewidth]{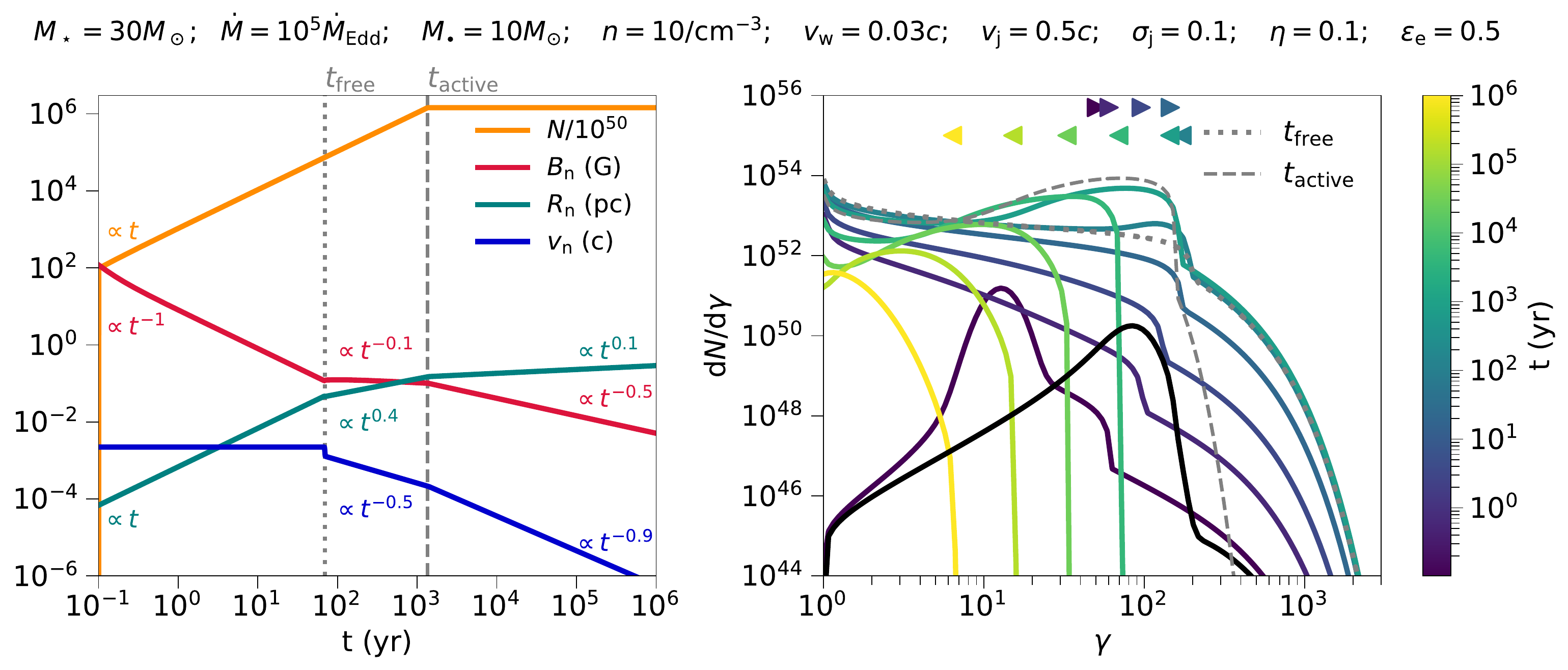}
\end{center}
\vspace{-0.25cm}
    \includegraphics[width=0.888\linewidth]{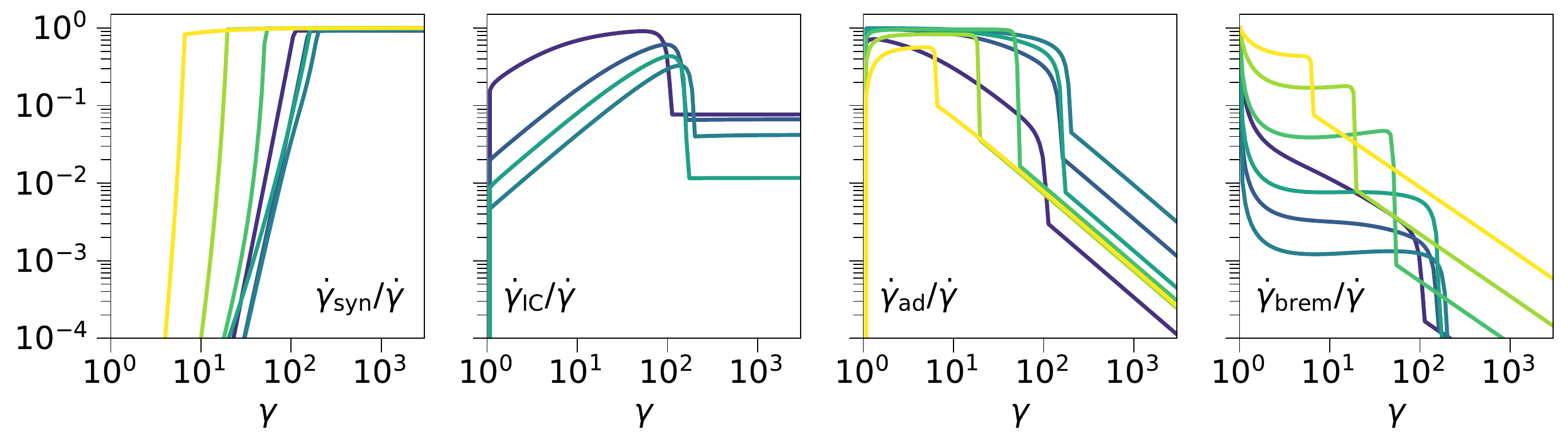}
\caption{Evolution of various intrinsic properties of the nebula (for the fiducial model with parameters listed in the figure title). Top-left panel: orange curves show the number of electrons in the nebula ($N/10^{50}$; Eq.~\ref{eq:continuity_eq}), red curve shows the magnetic field strength in the nebula ($B_{\rm n}$; Eq.~\ref{eq:B_n}), green curve shows the radius of the nebula ($R_{\rm n}$; Eq.~\ref{eq:R_n}), and the blue curve shows its expansion velocity ($v_{\rm n}={\rm d}R_{\rm n}/{\rm d}t$). The grey-dotted and -dashed vertical lines demarkate the critical times $t_{\rm free}$ (Eq.~\ref{eq:tfree}) and $t_{\rm active}$ (Eq.~\ref{eq:Mdot}). The approximate power-law temporal dependence of the parameters at different phases of the nebular evolution are also depicted with color-coding. Top-right panel: The energy distribution (${\rm d}N/{\rm d}\gamma = N_{\gamma}$) of the electrons in the nebula at different times (color-coded; see Eq.~\ref{eq:continuity_eq}). The distributions $N_\gamma$ at the critical times $t_{\rm free}$ and $t_{\rm active}$ are denoted by grey-dotted and -dashed curves, respectively. The color-coded markers at the top denote the electron Lorentz factor, $\gamma_{\rm syn}$, below which synchrotron cooling is inefficient on the nebula expansion time; the direction of the markers indicate the direction in which $\gamma_{\rm syn}$ evolves with time; the turnover happens at $t\sim t_{\rm free}$. The four panels in the bottom show the different components of electron energy losses included in our calculation (synchrotron, inverse-Compton, adiabatic, bremsstrahlung, from left to right; Eq.~\ref{eq:gdot}) for electrons with different energies $\gamma$, at different times (color-coded); note that $\dot{\gamma}_{\rm IC}/\dot{\gamma}\ll10^{-4}$ for $t>10^2$\,yr (not shown because it falls below the plotted range).
}
\label{fig:nebular_properties}
\end{figure}

The jet from the central accretion flow injects particles into the nebula with a luminosity $L_{\rm j}$ until $t=t_{\rm active}$ (Eqs.~\ref{eq:Mdot}, \ref{eq:injection_rate}). The left panel of Fig.~\ref{fig:nebular_properties} shows the total number of electrons in the nebula at different times; as expected, it stays constant for $t\ge t_{\rm active}$. The evolution of the nebular radius $R_{\rm n}$, its expansion velocity $v_{\rm n}$, and the internal magnetic field strength $B_{\rm n}$ are also shown, which follow the expected analytic relations (Eqs.~\ref{eq:R_n}, \ref{eq:B_n}). 

The right panel of Fig.~\ref{fig:nebular_properties} shows the electron energy distribution $N_\gamma$ at different times during the nebular evolution, obtained by numerically-integrating the continuity equation, Eq.~\eqref{eq:continuity_eq} considering various radiative losses (Eq.~\ref{eq:gdot}). The energy distribution $N_\gamma$ at the initial time follows that of relativistic Maxwellian of temperature $kT_{\rm in}=\bar{\gamma}_{\rm e}m_{\rm e}c^2\approx 60$ MeV ($\bar{\gamma}_{\rm e} \sim 100$; Eq.~\ref{eq:gammath}).  It soon rapidly evolves due to radiative losses that extends the electron population to lower energies. The electrons' energy spectra exhibit a `pile-up' bump at Lorentz factor $\gamma_{\rm syn}\gg10$.  Defining $\gamma_{\rm syn}$ as the electron Lorentz factor below which synchrotron losses are inefficient on the expansion time ($\dot{\gamma}_{\rm syn} < \dot{\gamma}_{\rm ad}$), we have marked its value with arrows along the top of Fig.~\ref{fig:nebular_properties}. The direction of the arrows denote the value of $\gamma_{\rm syn}$ initially increasing until $t\sim t_{\rm free}$, before turning over and beginning to decrease; this is also evident from the electrons' energy distribution at $t=t_{\rm free}$ (grey dotted curve), which is devoid of a bump at $\gamma_{\rm syn}$.  

The electrons are cooled at different times, depending on their energy, $\gamma$. We illustrate this in the bottom four panels of Fig.~\ref{fig:nebular_properties}: from left to right, each panel shows the fractional contribution of synchrotron ($\dot{\gamma}_{\rm syn}$), inverse-Compton ($\dot{\gamma}_{\rm IC}$), adiabatic ($\dot{\gamma}_{\rm ad}$), and bremsstrahlung ($\dot{\gamma}_{\rm brem}$) losses to the total energy loss rate ($\dot{\gamma}$; Eq.~\ref{eq:gdot}). At all times, the highest energy electrons (with $\gamma\gtrsim\gamma_{\rm syn}$) are cooled-down due to synchrotron losses. Lower energy electrons (with $\gamma \lesssim \gamma_{\rm syn}$) experience different dominant cooling processes at different phases of the nebular evolution: at earlier times $t<t_{\rm free}$, when the radiation density in the nebula is comparable to the magnetic energy, they are cooled down due to inverse-Compton scattering; at later times, $t>t_{\rm free}$, they are primarily coooled down due to adiabatic losses, with moderate losses via bremsstrahlung.  The fact that $\dot{\gamma}_{\rm syn}/\dot{\gamma}$ is at or near unity for the Lorentz factors of the injected thermal electron population $\bar{\gamma}_{\rm e} \sim 100$ shows that a large portion of the jet luminosity is radiated as synchrotron emission.

\begin{figure}
\centering
    \includegraphics[width=1\linewidth]{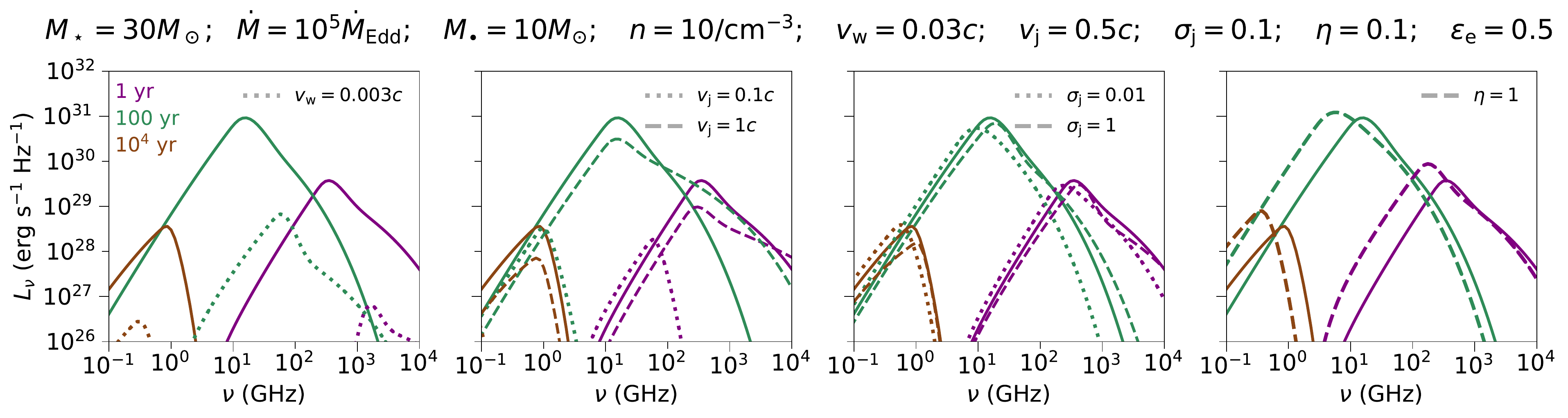}
\caption{Spectral energy distribution of the thermal synchrotron emission from the nebula. Each panel shows the spectrum at times 1\,yr (purple), 100\,yr (green), and $10^4$\,yr (brown). The fiducial model (whose parameters are listed in the figure title) is represented with solid curve, and the dotted and dashed curves in different panels represent changes to one of the parameters from the fiducial model ($v_{\rm w}$ in the first panel, $v_{\rm j}$ in the second panel, $\sigma_{\rm j}$ in the third panel, and $\eta$ in the fourth panel). In each panel, the fiducial model (whose parameters are listed in the figure title) is represented with solid curve, and the dotted and dashed curves in different panels represent changes to one of the parameters from the fiducial model ($v_{\rm w}$ in the first panel, $v_{\rm j}$ in the second panel, $\sigma_{\rm j}$ in the third panel, and $\eta$ in the fourth panel). 
}
\label{fig:Mdot1e5_spectra}
\end{figure}

\begin{figure} 
\centering
    \includegraphics[width=1\linewidth]{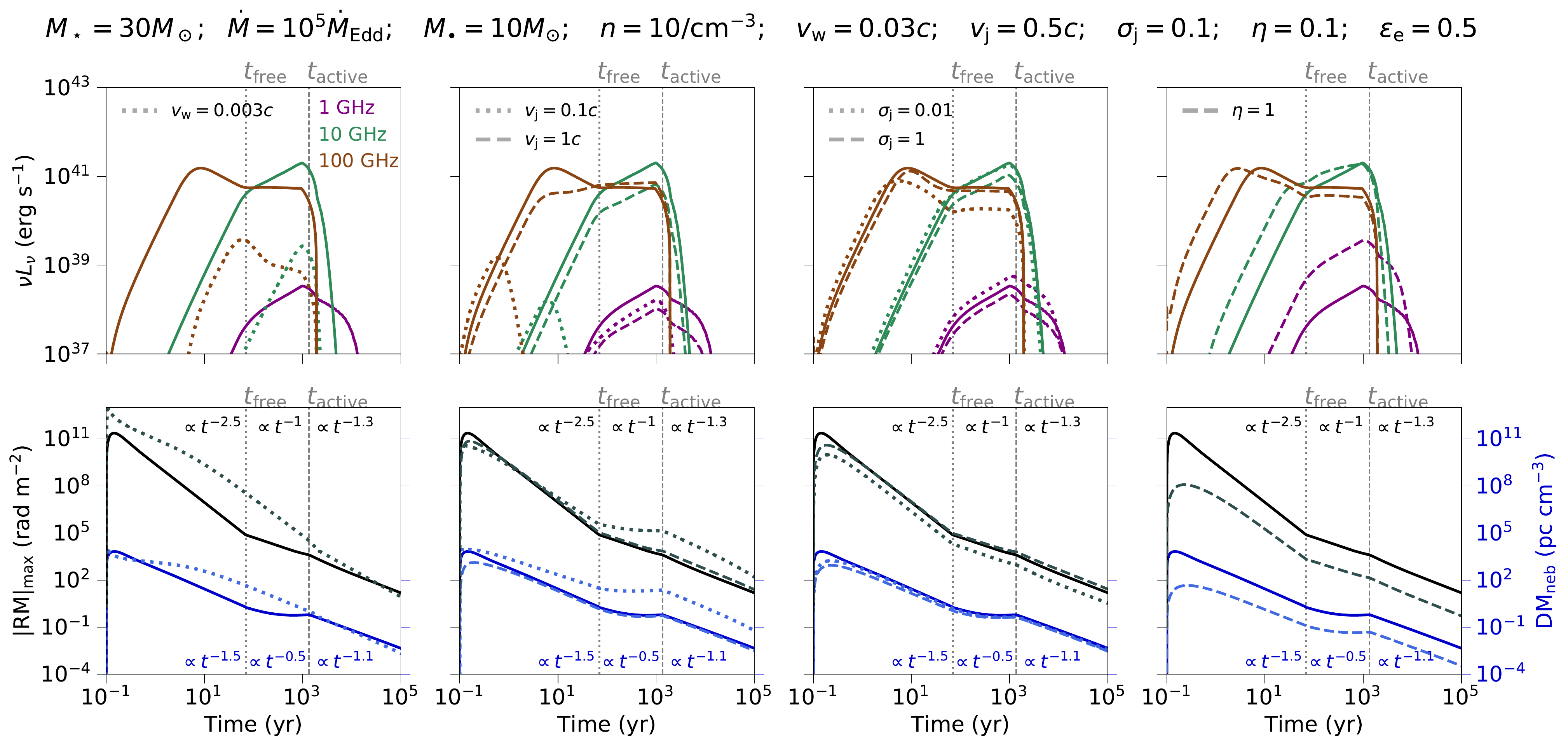}
\caption{Time evolution of observable parameters of the nebula for the same models shown in Fig.~\ref{fig:Mdot1e5_spectra}. Top panels show the light curve ($\nu L_{\nu}$) at frequencies 1\,GHz (purple), 10\,GHz (green), and 100\,GHz (brown); the bottom panels show the maximum rotation measure ($|$RM$|_{\rm max}$; black) and the local dispersion measure ($\text{DM}_{\rm neb}$; blue). The grey dashed and dotted vertical lines in each panel denote the active lifetime of the system ($t_{\rm active}$; Eq.~\ref{eq:Mdot}), and the transition timescale from free to self-similar expansion ($t_{\rm free}$; Eq.~\ref{eq:tfree}), respectively, for the fiducial model ($t_{\rm free}\approx2.2\times10^3$\,yr for the $v_{\rm w}=0.003\,c$ model). The temporal dependence of the fiducial $|$RM$|_{\rm max}$ (black) and $\text{DM}_{\rm neb}$ (blue) at different stages of the evolution is mentioned in the bottom panels. 
}
\label{fig:Mdot1e5_nuLnu_RM_DM}
\end{figure}

\subsubsection{Observable Properties} \label{subsubsec:observable_properties}

Fig.~\ref{fig:Mdot1e5_spectra} shows the spectral energy distribution at times 1\,yr (purple; $<t_{\rm free}$), 100\,yr (green; $t_{\rm free}<t<t_{\rm active}$), and $10^4$\,yr (brown; $>t_{\rm active}$). The spectrum peaks at the synchrotron frequency of the thermal electrons $\nu_{\rm pk}=\nu_{\rm syn, th}$ (Eq.~\ref{eq:nusynth}).  At this frequency SSA is typically negligible (i.e., $\tau(\gamma)=\alpha_\nu\nu_{\rm syn, th}R_{\rm n}<1$; Eq.~\ref{eq:f_ssa}). Likewise, free-free absorption  (Eq.~\ref{eq:ff_absorption}) through the wind shell---while relevant at very early times $t\ll t_{\rm free}$---is negligible during the light curve rise, with the free-free absorption frequency being well below the lowest range shown in Fig.~\ref{fig:Mdot1e5_spectra}.  

While $\nu_{\rm pk}$ evolves to lower frequencies with time, the peak luminosity $L_\nu(\nu_{\rm pk})$ increases until $t\sim t_{\rm active}$, after which it decreases sharply once the outflow has turned off.  The product $\nu_{\rm pk} L_\nu(\nu_{\rm pk})$ remains roughly constant in time, as expected in the regime of efficient synchrotron cooling (Eq.~\ref{eq:syncool_exp_timescale}), for which the bolometric luminosity approximately tracks the (temporally constant) rate of injected electron energy $\sim \varepsilon_{\rm e} L_{\rm j}$.  For greater accretion/jet efficiency $\eta$, $L_\nu(\nu_{\rm pk})$ mildly increases and $\nu_{\rm pk}$ decreases.  Changes to $\sigma_{\rm j}$ barely affects $\nu_{\rm pk}$ and $L_\nu(\nu_{\rm pk})$. As far as the dependence on the jet speed $v_{\rm j}$, $L_\nu(\nu_{\rm pk})$ reaches a maximum for $v_{\rm j}\sim 0.5\,c$; the peak frequency $\nu_{\rm pk}$, however, is much larger for smaller $v_{\rm j} \sim0.1 \,c$. At early times $t\lesssim t_{\rm active}$, larger values of $v_{\rm w}$ lead to a decrease in $\nu_{\rm pk}$ and an increase in $L_\nu(\nu_{\rm pk})$. On the other hand, at late times $t\gtrsim t_{\rm active}$, larger values of $v_{\rm w}$ increase both $\nu_{\rm pk}$ as well as $L_\nu(\nu_{\rm pk})$.

The top row of Fig.~\ref{fig:Mdot1e5_nuLnu_RM_DM} shows the light curve in three different bands (1\,GHz: purple, 10\,GHz: green, and 100\,GHz: brown).  The highest frequency light curves peak as the spectral peak crosses down through that bandpass (Fig.~\ref{fig:Mdot1e5_spectra}), while at lower frequencies the spectral peak never reaches the observing band by $t_{\rm active}$ so the light curve continues to rise until $t \sim t_{\rm active},$ before falling off sharply thereafter.  The peak luminosity of the fiducial model reached at $t=t_{\rm pk}$ is, $\nu L_\nu(t_{\rm pk})\gtrsim10^{41}$\,erg\,s$^{-1}$ for 100\,GHz and 10\,GHz, but is much smaller $\lesssim10^{39}$\,erg\,s$^{-1}$ for 1\,GHz. The peak luminosity decreases significantly by a factor ${\cal O}(10^2)$ for either a smaller $v_{\rm w}=0.003\,c$ or $v_{\rm j}=0.1\,c$ compared to the fiducial values, but are nearly independent of $\sigma_{\rm j}$. Furthermore, a smaller value of $v_{\rm w}$ ($v_{\rm j}$) delays (advances) the peak time $t_{\rm pk}$ achieved in all higher frequency bands ($\gg1$\,GHz).  While the peak luminosity $\nu L_\nu(t_{\rm pk})$ at high frequencies (10-100\,GHz) is relatively independent of the value of $\eta$, the $\nu L_\nu(t_{\rm pk})$ at 1\,GHz decreases from $\sim10^{40}$\,erg\,s$^{-1}$ to $\sim10^{38}$\,erg\,s$^{-1}$ as $\eta$ decreases from 1 to 0.1.  

Defining the duration of the broad peak as the time span $\Delta t_{\rm pk}$ over which $\nu L_\nu > 50\%\nu L_\nu(t_{\rm pk})$, we find $\Delta t_{\rm pk}={\cal O}(10^2)$\,yr at 1\,GHz and 10\,GHz in the fiducial model, which decreases to ${\cal O}(1)$\,yr at 100\,GHz. The peak duration $\Delta t_{\rm pk}$ also decreases for smaller $v_{\rm j}\sim0.1\,c$ to $\Delta t_{\rm pk}\sim{\cal O}(1)$\,yr at 100\,GHz and $\Delta t_{\rm pk}\sim{\cal O}(10)$\,yr at 10\,GHz; $\Delta t_{\rm pk}$ at lower frequencies $\sim1$\,GHz is insensitive to changes in $v_{\rm j}$. On the other hand, changes to $v_{\rm w}$ or $\sigma_{\rm j}$ or $\eta$ only modestly changes the peak duration.

The bottom row of Fig.~\ref{fig:Mdot1e5_nuLnu_RM_DM} shows the temporal evolution of $|$RM$|_{\rm max}$ and $\text{DM}_{\rm neb}$. For our fiducial model, $|$RM$|_{\rm max}$ starts at a very high value but decreases monotonically, as $\propto t^{-2.5}$ during $t<t_{\rm free}$, $|$RM$|_{\rm max}\propto t^{-1}$ during $t_{\rm free}<t<t_{\rm active}$, and $|$RM$|_{\rm max}\propto t^{-1.3}$ during $t>t_{\rm active}$. This trend is preserved for changes in $\sigma_{\rm j}$ and $\eta$, but with a decreasing normalization for smaller $\sigma_{\rm j}$ or $\eta$. On the other hand, a decrease in $v_{\rm w}$ or $v_{\rm j}$ leads to an increase in the $|$RM$|_{\rm max}$, with a mildly shallower slope.  As expected, the evolution of the nebular contribution to the DM behaves qualitatively similar to the $|$RM$|_{\rm max}$ for similar variations to the system parameters, however with a rather shallower decrease. E.g., for our fiducial model, $\text{DM}_{\rm neb}\propto t^{-1.5}$ during $t<t_{\rm free}$, $\text{DM}_{\rm neb}\propto t^{-0.5}$ during $t_{\rm free}<t<t_{\rm active}$, and $\text{DM}_{\rm neb}\propto t^{-1.1}$ during $t>t_{\rm active}$. Overall, we see that the contribution of the nebula to the DM is comparable to that of the entire shell's contribution (Eq.~\ref{eq:DM}). We note here that for systems with much longer $t_{\rm active}$ (or a smaller $t_{\rm free}$), the $|$RM$|_{\rm max}$ and $\text{DM}_{\rm neb}$ tend to exhibit a mildly increasing trend for $t\gg t_{\rm free}$ (see Eqs.~\ref{eq:DM}, \ref{eq:RM_analytical}).

We conclude with the reminder that the light curves and spectra shown in Fig.~\ref{fig:Mdot1e5_spectra} are calculated assuming a one-zone nebula with the injected electrons possessing a single temperature.  Though a useful idealization, the actual nebula is likely to be inhomogeneous and seeded with electrons with a range of different energies, some even belonging to a non-thermal population; these effects will broaden the spectrum around $\nu_{\rm pk}$ considerably relative to the model predictions (we return to this issue when fitting individual sources in Sec.~\ref{subsec:frb_application}). 

\subsection{Dependence on Mass-Transfer Rate} \label{subsec:parameter_study}

\begin{figure}
\centering
    \includegraphics[width=1\linewidth]{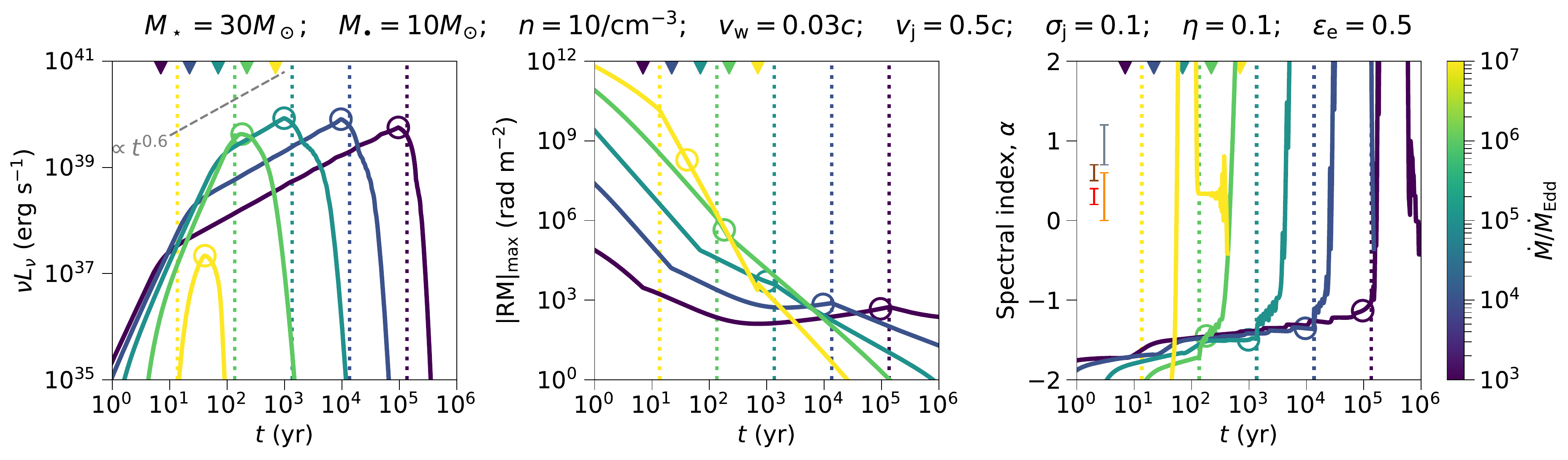}
\caption{Left panel: 3\,GHz radio light curves; the grey-dashed line is representative of the $\propto t^{0.6}$ time dependence that the light curves show. Middle panel: time evolution of $|$RM$|_{\rm max}$. Right panel: 2--4\,GHz spectral index $\alpha$, where the flux density, $S_{\nu}\propto\nu^{-\alpha}$; the vertical error bars---barring their location along the time axis---are representative range of spectral indices for various astronomical sources (red: FRBs, brown: GRBs, orange: pulsar wind nebul\ae, grey: supernov\ae\ and tidal disruption events). In all the panels, the color of the curves denote different $\dot{M}/\dot{M}_{\rm Edd}$, with the other fiducial parameters listed in the figure title; dotted vertical lines and the downward pointing triangles denote the critical times $t_{\rm active}$ (Eq.~\ref{eq:Mdot}) and $t_{\rm free}$ (Eq.~\ref{eq:tfree}), respectively, for each model. The colored circles denote the parameter value at $t_{\rm pk}$.}
\label{fig:Mdot_lc_RM_index}
\end{figure}

\begin{figure}
\centering
    \includegraphics[width=1\linewidth]{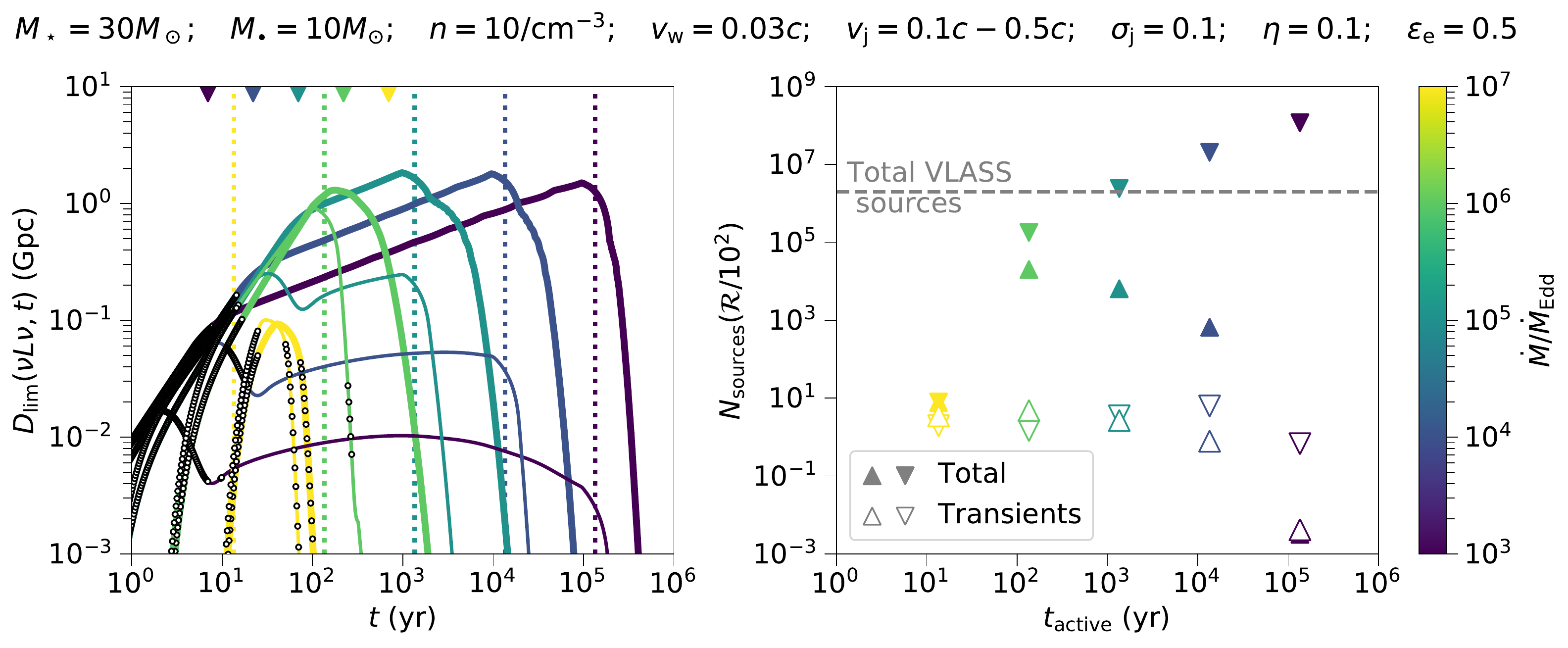}
\caption{Left panel: Radio (3\,GHz) detection horizon (Eq.~\ref{eq:detection_horizon}). Thin and thick curves correspond to models with $v_{\rm j}=0.1\,c$ and $v_{\rm j}=0.5\,c$, respectively; dotted vertical lines and the downward pointing triangles denote the critical times $t_{\rm active}$ (Eq.~\ref{eq:Mdot}) and $t_{\rm free}$ (Eq.~\ref{eq:tfree}), respectively, for each model. The black circles scattered along the curves denote the times when the source will be detected as a transient by VLASS (see Sec.~\ref{subsec:survey_detection} for the criterion). Right panel: Number of detectable radio (3\,GHz) sources as a function of the active duration of the system ($t_{\rm active}$; Eq.~\ref{eq:Mdot}), considering separately the total number of sources (Eq.~\ref{eq:N_tot}; solid markers) and just those that will be detected as transients in VLASS (Eq.~\ref{eq:N_transient}; hollow markers), for an assumed rate $\mathcal{R} = 100$ Gpc$^{-3}$ yr$^{-1}$.  The upward-and downward-pointing triangle markers denote models with $v_{\rm j}=0.1\,c$ and $v_{\rm j}=0.5\,c$, respectively. The calculations are performed with parameters (listed in the figure title); the color-coding denotes different $\dot{M}/\dot{M}_{\rm Edd}$.  
}
\label{fig:Mdot_detectability}
\end{figure}

Motivated by the wide range of possible mass-transfer rates in different binary systems in different evolutionary states (Sec.~\ref{subsec:binary}), we now explore how the nebular observables vary for higher and lower values of $\dot{M}$ (or, equivalently for fixed $M_{\star}$, different active periods $t_{\rm active} \propto \dot{M}^{-1}$), keeping the other fiducial model parameters fixed from those assumed in Sec.~\ref{sec:singleMdot}.  We focus on an observing frequency around 3 GHz, matched to Very Large Array Sky Survey (VLASS, \citealt{Lacy+20}; see Sec.~\ref{subsec:survey_detection}).

In the left panel of Fig.~\ref{fig:Mdot_lc_RM_index}, we show the 3\,GHz light curve color-coded by $\dot{M}$. For moderately accreting systems ($10\lesssim\dot{M}/\dot{M}_{\rm Edd}\lesssim10^5$) with $t_{\rm active}>t_{\rm free}$, the light curve rises as a power-law $\nu L_\nu \propto t^{0.6}$ at times $t_{\rm free}<t<t_{\rm active}$.  Except in the highest $\dot{M} = 10^{7}\dot{M}_{\rm Edd}$ case, the 3 GHz peak is achieved at $t_{\rm pk} \sim t_{\rm active} \propto\dot{M}^{-1}$ (colored dotted vertical lines).  The 3 GHz peak luminosities are remarkably stable at $\nu L_\nu\sim10^{40}$\,erg\,s$^{-1}$ for wide a range of $\dot{M}/\dot{M}_{\rm Edd}\lesssim10^6$; this is in contrast to the peak bolometric luminosity, which we find increases $L_{\rm bol} \propto\dot{M}^{0.5}$. The fact that $L_{\rm bol}$ does not increase precisely in proportion to the injected electron power $\propto L_{\rm j} \propto \dot{M}$ likely results from synchrotron cooling not being as efficient as predicted by analytical estimates (e.g., Eq.~\ref{eq:syncool_exp_timescale}), due to suppression of the electron cooling rate by SSA at $\nu_{\rm syn,th}$ (Eq.~\ref{eq:f_ssa}). By contrast, for the highest accretion rate systems $\dot{M}/\dot{M}_{\rm Edd}\gtrsim 10^{6}$ with the shortest active lifetimes $t_{\rm active}< t_{\rm free}$, the peak luminosity is considerably lower $\nu L_\nu\ll10^{40}$\,erg\,s$^{-1}$ and peaks after the engine has turned off ($t_{\rm pk}>t_{\rm active}$).  

The similar peak luminosity and temporal rise-rate attained for a wide range of $\dot{M}/\dot{M}_{\rm Edd}\lesssim10^6$ could make it challenging to obtain meaningful constraints on the systems' properties (e.g., age, $\dot{M}$, etc.) given just a relatively short ($\sim$years- to decades-long) time-span light curve around 3\,GHz.  Fortunately, other observable properties such as the location of the spectral peak (e.g., high-frequency radio observations with VLA or ALMA) and the maximum rotation measure (right panel of Fig.~\ref{fig:Mdot_lc_RM_index}) depend more sensitively on $\dot{M}$ at a given system age.  For all $\dot{M}$, we see that $|$RM$|_{\rm max}$ follows $\propto t^{-2}$ at early times $t<t_{\rm free}$, and its normalization scales with $L_{\rm j} \propto \dot{M}$, in rough agreement with Eq.~\eqref{eq:RM_analytical}. However, it is only the moderately accreting systems ($10\lesssim\dot{M}/\dot{M}_{\rm Edd}<10^5$)---with sufficiently long $t_{\rm active}$ allowing for expansion---that exhibit a phase of increasing $|$RM$|_{\rm max}$ ($\propto t^{0.7}$ for $t_{\rm free} < t < t_{\rm active}$), as predicted in Eq.~\eqref{eq:RM_analytical}. At times $t>t_{\rm active}$, $|$RM$|_{\rm max}$ decreases again for all models, albeit at a faster rate for higher $\dot{M}$ systems. 

\section{Detection Prospects} \label{sec:discussion}

We now discuss different methods to detect, identify, and characterize ULX hyper-nebul\ae, and then discuss on the applicability of our model to FRB persistent radio sources and local ULX bubbles.

\subsection{Detection in Blind Surveys} \label{subsec:survey_detection}

Based on population synthesis modeling, \citet{Schroder+20} estimate that NS/BH common envelope events with donor stars more massive than 10$M_{\odot}$ occur at a frequency of $\sim 0.6\%$ the core collapse supernovae rate, corresponding to a volumetric rate in the local universe of ${\cal R} \sim 10^{3}$ Gpc$^{-3}$ yr$^{-1}$ (\citealt{Strolger+15}; see also \citealt{Vigna-Gomez+18}).
\citet{Pavlovskii+17} estimate a Milky Way formation
rate of $\sim 3\times 10^{-5}$ yr$^{-1}$ for massive star binaries with mass-transfer rates $\dot{M} \gtrsim 100\dot{M}_{\rm Edd},$ corresponding to ${\cal R} \sim 10^{2}$ Gpc$^{-3}$ yr$^{-1}$.  The rate of engine-powered FBOTs  (${\cal R} \sim 2-400$ Gpc$^{-3}$ yr$^{-1}$; \citealt{Coppejans+20,Ho+21b}), which may be associated with failed common envelope events \citep{Soker+19,Schroder+20,Metzger22}, is consistent with these estimates, but could be more than an order of magnitude lower.

One method to discover ULX hyper-nebula\ae~is with wide-field blind radio surveys, such as VLASS \citep{Lacy+20} or by comparing VLASS with earlier survey data such as FIRST \citep{Becker+95}.  Given a peak luminosity $\nu L_{\nu}$ and peak duration $\Delta t_{\rm pk}$, the total number of sources at any time across the entire sky above a given flux density $F_{\rm lim}$ can be estimated by
\be \label{eq:N_tot}
N_{\rm tot} \simeq \frac{4\pi}{3}D_{\rm lim}(\nu L_\nu, t_{\rm pk})^{3}{\cal R}\Delta t_{\rm pk}. 
\ee
where ${\cal R}$ is the source volumetric rate and 
\be \label{eq:detection_horizon}
 D_{\rm lim} = \left(\frac{L_\nu}{4\pi F_{\rm lim}}\right)^{1/2} \simeq 0.33\,{\rm Gpc}\,\left(\frac{L_{\nu}}{10^{29}\,{\rm erg\,s^{-1}\,Hz^{-1}}}\right)^{1/2}\left(\frac{F_{\rm lim}}{0.7\,{\rm mJy}}\right)^{-1/2}
\ee
is the detection horizon for a limiting flux sensitivity, $F_{\rm lim}$, neglecting cosmological effects on the luminosity distance and source evolution.  We have normalized $F_{\rm lim}$ to a value of 0.7 mJy, corresponding to the $5\sigma$ flux sensitivity of the VLASS around 3 GHz \citep{Lacy+20}.  The form of Eq.~\eqref{eq:detection_horizon} assumes that $N_{\rm tot}$ is dominated by sources near their peak luminosity; this is typically justified because $\nu L_{\nu}$ after peak falls off faster than $\propto 1/t$---especially at lower frequencies $\nu\sim3$\,GHz (see Fig.~\ref{fig:Mdot1e5_nuLnu_RM_DM}).  

The colored curves in the left panel of Fig.~\ref{fig:Mdot_detectability} shows $D_{\rm lim}$ as a function of time for the suite of light curve models from Fig.~\ref{fig:Mdot_lc_RM_index}, as well as a second set of light curve models calculated assuming otherwise identical parameters, but with a lower $v_{\rm j} = 0.1\,c$ (equivalently, for mean injected electron Lorentz factor $\sim 25$ times lower than the $v_{\rm j} = 0.5\,c$ model; Eq.~\ref{eq:gammath}). The right panel shows $N_{\rm tot}(F_{\rm lim} = 0.7$\,mJy) as a function of $t_{\rm active}$, separately for the $v_{\rm j} = 0.1\,c$ and $v_{\rm j} = 0.5\,c$ models, normalized to a total rate per $t_{\rm active}-$bin of ${\cal R} = 10^{2}$\,Gpc$^{-3}$\,yr$^{-1}$, i.e. about 10\% of the total NS/BH common envelope event rate \citep{Schroder+20}.  The predicted number of ULX hyper-nebul\ae~in VLASS is extremely sensitive to the model assumptions, increasing from $\lesssim 0.1-10\%$ of the total VLASS source count ($N_{\rm VLASS} \sim 2\times 10^{6}$) for the $v_{\rm j} = 0.1\,c$ models to $\gtrsim 100\%$ of $N_{\rm VLASS}$ for $v_{\rm j} = 0.5\,c$.  Such an over-production should be interpreted as meaning the formation rate of systems with the parameters of our $v_{\rm j} = 0.5\,c$ model is constrained to be far less than ${\cal R} \sim 10^{2}$ Gpc$^{-3}$ yr$^{-1}$ (indeed, we find in Sec.~\ref{subsec:frb_application} that FRB persistent sources can be fit to a model with $v_{\rm j} = 0.2\,c$).

A detection in VLASS may not be sufficient to uniquely identify ULX hyper-nebul\ae, given the many other extragalactic radio source populations.  However, time-evolution of the source flux over multiple epochs may provide a distinctive diagnostic, particularly for young sources.  We estimate the number of sources detected as {\it transient} in VLASS, 
\be \label{eq:N_transient}
N_{\rm transient} \simeq \sum_{i} \frac{4\pi}{3}D_{\rm lim}(\nu L_{\nu,i},t_{i})^{3}{\cal R}\Delta t_{\rm i},
\ee
where now the summation is performed by counting those time intervals $\Delta t_{\rm i}$ over which $\nu L_\nu$ changes by a factor of $\ge 3$ within a span of 20\,yr.  We choose this timescale as a reference to the time gap between the FIRST and VLASS survey, similar to the criterion used by \citet{Law+18} and \citet{Dong+21} to identify radio transients (although other criterion can be adopted e.g., change in $\nu L_\nu$ by a factor of 50\% over 4\,yr for a source to be detected as a transient within VLASS epochs, which we defer to future work).  The epochs which satisfy our adopted variability criterion are shown with black dots in the left panel of Fig.~\ref{fig:Mdot_detectability}, while the detectable number of transients $N_{\rm transients}$ are shown as open triangles in the right panel.  We find that only a handful $N_{\rm transient} \sim 1-10$ of the total detected sources are identifiable as transients according to the adopted criterion.  These are mostly systems detected early in the light curve rise, when the source is still young $t \lesssim 10^{2}$ yrs.  Since their luminosities have not yet peaked, these ``transients'' are located at nearby distances $D_{\rm lim} \sim 100$ Mpc compared to the total source population ($D_{\rm lim} \sim 1$ Gpc). 

As a result of the nearby distances of the VLASS transients, follow-up observations at other electromagnetic wavelengths may be fruitful for these sources.  Unfortunately though, the X-ray luminosity of the nebula-CSM shock at such early times $\lesssim 10^{2}$ yr is too modest $L_{\rm X} \lesssim 10^{37}$ erg s$^{-1}$ (Fig.~\ref{fig:L_X}) to be detected with current X-ray telescopes, even at tens of Mpc.  X-ray emission from the inner accretion flow, similar to that which give ULX their namesake, would be considerably more luminous; however, due to geometric-beaming of the emission by the accretion funnel, which may become even narrower at such high accretion rates, this emission is likely only observable for a small fraction of viewing angles (Sec.~\ref{subsec:frb_application}).  The nearest non-transient ULX hyper-nebul\ae~could be considerably older and hence more X-ray luminous, making them promising targets for future large X-ray observatories like \textit{Athena} \citep{Barcons+17} or \textit{AXIS} \citep{Mushotzky+19}.  However, absent the rapid time-variability, additional characteristics such as the radio spectral properties (Fig.~\ref{fig:Mdot_lc_RM_index}), or host galaxy demographics \citep{Kovlakas+20,Sridhar+21b}, may be required to identify a candidate sample. 
As discussed in Sec.~\ref{subsubsec:X-ray}, the line cooling contribution of the shock-ionized plasma of the hyper-nebul\ae~would predominantly be emitted in UV/optical bands. From Fig.~\ref{fig:L_X}, and Eqs.~\eqref{eq:cooling_function} and \eqref{eq:L_X}, we predict---for optimistic scenarios---a peak optical counterpart to the hyper-nebul\ae\ with a luminosity of $10^{39}$\,erg\,s$^{-1}$, corresponding to a V-band apparent magnitude of $\sim24$ for sources located at $\sim100$\,Mpc. In addition, reprocessing of beamed X-rays from the ULX-jet can also contribute the optical/UV emission \citep{Pakull&Mirioni02, Wang02, Ramsey+06, Soria+10, Sridhar+21b}.  Given the relatively smaller sizes ($\sim$\,pc) and brighter nature of the hyper-nebul\ae~than their older, long-lived ULX-nebul\ae\ counterparts ($\sim10^2$\,pc), these sources will likely be unresolved, and if sufficiently close, may be detected in optical surveys such as Pan-STARRS \citep{Tonry+12}, Zwicky Transient Facility \citep{Masci+19, Graham+19, Bellm+19} and most likely with the upcoming Vera C. Rubin Observatory \citep{Ivezic+19}. We recommend follow-up observations of hyper-nebul\ae~candidates detected with optical/radio surveys with Hubble Space Telescope (HST), Dark Energy Camera \citep[DECam;][]{Flaugher+15}, and Very-Long-Baseline Interferometry.

\subsection{Application to FRB Persistent Radio Sources} \label{subsec:frb_application}

\begin{figure} 
\centering
    \includegraphics[width=1\linewidth]{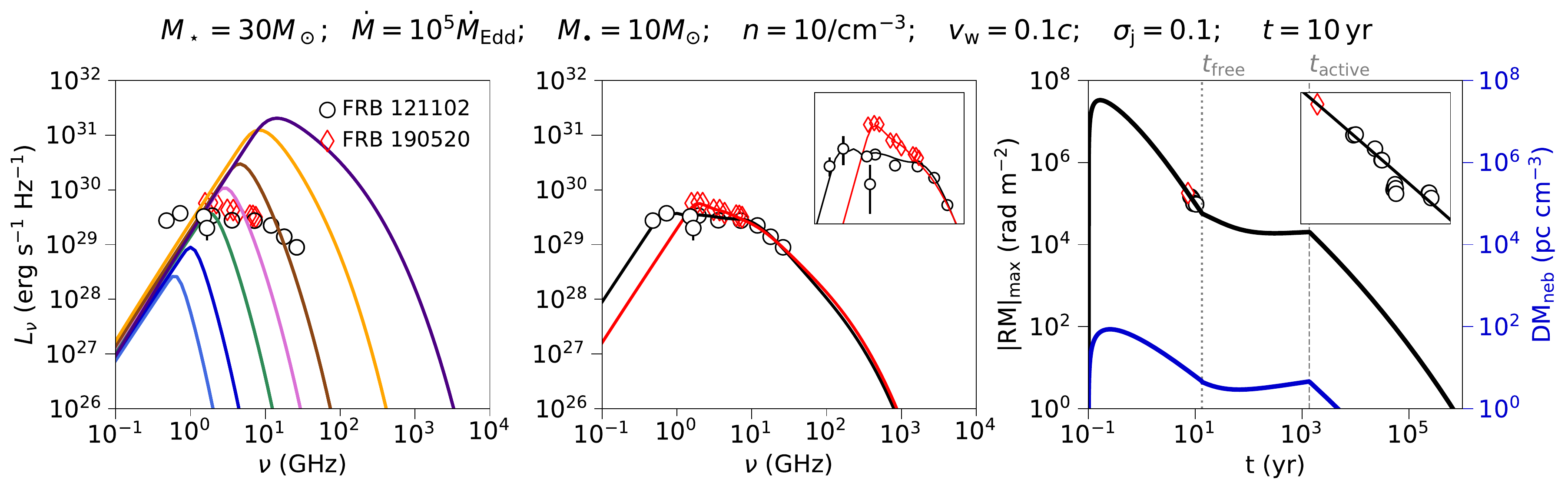}
\caption{Application of the model for disk-wind inflated ULX radio hyper-nebul\ae\ to FRB sources. The left panel shows the observed spectra of the persistent radio source counterparts to \prsone~(black circles; observed with VLA in 1.6-22\,GHz range by \citealt{Chatterjee+17} and with GMRT in 0.4-1.4\,GHz range by \citealt{Resmi+21}), and \prstwo~(red diamonds; observed with FAST in 1.3-6\,GHz range by \citealt{Niu+22}), respectively. The solid curves are the synchrotron radio spectra calculated for a grid of models with parameters listed in the figure title, with the following differences: going from blue (left) to indigo (right) curves, $v_{\rm j}$ increases from $0.15\,c - 0.4\,c$, $\varepsilon_{\rm e}$ increases from $0.15 - 0.8$, and $\eta$ decreases from $1 - 0.4$ (such that the jet luminosity $L_{\rm j} \propto \eta v_{\rm j}^{2}$ remains roughly fixed).  Middle panel shows a fit to the observed data with models (red solid curve for \prstwo~ and black solid curve for \prsone) obtained by performing a weighted-sum of the grid of models in the left panel. The right panel shows the $|$RM$|_{\rm max}$ of \prsone~(black circles; \citealt{Michilli+18, Hilmarsson+21}) and \prstwo~(red diamond; \citealt{Niu+22,Anna-Thomas+22,Feng+22}) at different times. The black and blue solid curves denote the local nebular contribution to the $|$RM$|$ and DM calculated for the model corresponding to the green spectral curve in the left panel (more details in the text, Sec.~\ref{subsec:frb_application}). The vertical grey dotted and dashed lines denote the ejecta free expansion timescale ($t_{\rm free}$; Eq.~\ref{eq:tfree}) and the active timescales of the system ($t_{\rm active}$; Eq.~\ref{eq:Mdot}), respectively. The insets in the middle and the right panels offer a zoomed-in view of the region around the observed data.}
\label{fig:FRB_fit}
\end{figure}

ULX nebul\ae~may also serve as signposts to repeating FRB sources, in models where the bursts arise from shocks or reconnection events in the accretion-powered jet \citep{Sridhar+21b}.  The PRSs spatially coincident with the engines of repeating FRB are compact sources of incoherent synchrotron radiation that are too luminous ($L_\nu \gtrsim 10^{29}$\,erg\,s$^{-1}$\,Hz$^{-1}$) to be explained by star formation activity in the host galaxy, or individual supernova remnants (e.g., \citealt{Law+22}).  A transient radio source, with a luminosity and spectral properties similar to FRB PRS, was recently discovered in VLASS \citep{Dong+22}. In this section, we apply our model of ULX hyper-nebul\ae\ to explain the observed properties of some of PRSs from some of the brightest FRBs (see Sec.~\ref{subsec:local_ULX} for more discussion on our model's relevance to fainter FRBs and XRBs).

The radio spectra of observed PRS are relatively flat and typically broader than those predicted by our one-zone model near its peak $\nu_{\rm pk}$ (see Fig.~\ref{fig:Mdot1e5_spectra} and right panel of Fig.~\ref{fig:Mdot_lc_RM_index}).  However, the nebula model we have presented thus far is overly idealized: it assumes all of the electrons injected into the nebula possess a single temperature, as set by a single velocity of the jet termination shock $v_{\rm j}$ and a single electron heating efficiency $\varepsilon_{\rm e}$.  In reality, due to time-variation in the jet speed and/or spatial variation (e.g. curvature or angle-dependent jet velocity) across the termination shock, the nebula will very likely be seeded with electrons containing a range of different temperatures and corresponding mean energies $\bar{\gamma}_{\rm e} \propto \varepsilon_{\rm e} v_{\rm j}^{2}$ (Eq.~\ref{eq:gammath}); the effect of a width to the $\bar{\gamma}_{\rm e}$ distribution will be to broaden the spectrum around $\nu_{\rm pk}$ relative to the baseline model.  Consider that even a relatively modest, factor $\sim 2$ variation in the jet shock velocity causes a factor $\sim 4$ spread in the mean electron energy $\bar{\gamma}_{\rm e} \propto v_{\rm j}^{2}$ (Eq.~\ref{eq:gammath}) and hence $\sim 16$ in the peak radio frequency $\nu_{\rm syn,th} \propto \bar{\gamma}_{\rm e}^{2}$ (Eq.~\ref{eq:nusynthnum}).

Fig.~\ref{fig:FRB_fit} shows fits of our model to the PRSs associated with \prsone~and \prstwo, where we have now relaxed the assumption of a single value of $v_{\rm j}$ and $\varepsilon_{\rm e}$.   The left panel of Fig.~\ref{fig:FRB_fit} shows the observed PRS spectra, and a grid of model spectral curves calculated for a ULX hyper-nebula with the following parameters: $M_\star=30M_{\odot}$, $\dot{M}=10^5\dot{M}_{\rm Edd}$, $M_\bullet=10M_\odot$, $n=10$\,cm$^{-3}$, $v_{\rm w}=0.1\,c$, $\sigma_{\rm j}=0.1$, at a time $t=10$\,yr; different spectral curves are produced by varying the following parameters within the following ranges: $v_{\rm j}/c\in[0.15,0.4]$, $\varepsilon_{\rm e}\in[0.15,0.8]$, and $\eta\in[0.4,1]$ (for \prstwo, we vary the parameters within the restrictive range $v_{\rm j}/c\in[0.2,0.4]$ and $\varepsilon_{\rm e}\in[0.2,0.8]$).  We vary these three parameters such that\footnote{Although this precise dependence is arbitrary, particle-in-cell simulations predict the electron heating efficiency to increase with increasing shock speed moving from the non-relativistic regime \citep{Tran&Sironi20} to the relativistic regime \citep{Sironi&Spitkovsky11}.} $\varepsilon_{\rm e}\propto v_{\rm j}$, at a fixed jet luminosity $L_{\rm j} \propto \eta v_{\rm j}^2 = constant$.  In order to reproduce the observed PRS spectra, we have generated a weighted-sum of the spectral curves from these models.  The specific choice of the weights $w$, is such that $w\propto \bar{\gamma}_{\rm e}^{-2} \propto (\varepsilon_{\rm e}v_{\rm j}^2)^{-2}$, with minor modifications around it to fit spectra from different PRSs.  This general approach of a linear weighting of the models is justified by the fact that electrons injected into the same nebula with different energies (e.g., at different epochs if $v_{\rm j}$ is varying in time, or at different locations if the termination shock properties vary spatially) will evolve and radiate largely independently of each other.  The final weighted-spectra are shown in the middle panel of Fig.~\ref{fig:FRB_fit}, revealing a satisfactory (if admittedly tuned) proof-of-principle match to \prsone~and \prstwo.

The local contributions to the RM and DM will arise primarily from the region of the nebula that is energetically dominant. We therefore choose the model (green curve in left panel of Fig.~\ref{fig:FRB_fit}) with the largest value of $w\varepsilon_{\rm e}v_{\rm j}^2$, and show the $|$RM$|_{\rm max}$ and $\text{DM}_{\rm neb}$ of this model in the right panel. \cite{Hilmarsson+21} report that the RM of \prsone~has decreased from $1.46 \times 10^5$\,rad\,m$^{-2}$ in 2017 January to $9.7 \times 10^4$\,rad\,m$^{-2}$ by 2019 August (i.e., a decrease of $\sim$15\% yr$^{-1}$). Our model reproduces this RM decay (top-right inset), and we retrieve an age of $\sim$10\,yr for \prsone, consistent with the age assumed in our spectral fit (see also \citealt{Margalit&Metzger18}). Furthermore, the $|$RM$|_{\rm max}$ measured for \prstwo~at $>1.8\times10^5$\,rad\,m$^{-2}$ (measured assuming 100\% linear polarization; \citealt{Niu+22}) is also captured by the same model that describes the RM of \prsone, consistent with \prstwo~being a younger version of \prsone.  This justifies using the same spectral model to fit both \prsone~and \prstwo, with changes only in the weight factors. However, we note that the analysis of \cite{Zhao&Wang_21}---assuming a different external medium (composite of magnetar wind nebula and supernova remnant)---estimated the age of \prstwo~(16--20\,yr) to be older than \prsone~(14\,yr).  Unlike \citet{Margalit&Metzger18} and \cite{Zhao&Wang_21}, our model can reproduce the observations even for a temporally-constant injected power $L_{\rm j}$. 

Recent observations \citep{Anna-Thomas+22, Dai+22} have revealed rapid changes to the RM of \prstwo~($\sim$300\,rad\,m$^{-2}$\,${\rm day}^{-1}$), as well as a `zero-crossing' reversal.  For the periodic repeating source FRB 20180916B \citet{Mckinven+22} found an increase in the RM of of over 40\% over only a 9-month interval, which appears unrelated to the $\approx 16.3$ day activity cycle phase.  As discussed near the end of Sec.~\ref{subsubsec:analytic_sync_RM}, our model predicts only the long-term secular evolution of $|$RM$|_{\rm max}$, and not of the detailed magnetic field structure in the nebula which is also necessary to determine RM.   More rapid, potentially stochastic variations (including possible sign-reversals) of magnitude $|\dot{{\rm RM}}| \sim |$RM$|_{\rm max}/t_{\delta \rm RM}$ are possible on timescales as short as the shock's dynamical timescale ($t_{\delta \rm RM} \sim$ weeks to months for the $\sim$decades-old sources of interest; Eq.~\ref{eq:t_deltaRM}). Such fluctuations can arise due to turbulent motions in the nebula \citep{Feng+22,Katz+22, Yang+22}, for instance generated by vorticity or magnetic dissipation ahead of the jet termination shock (Fig.~\ref{fig:cartoon}).  Similar turbulence may induce scatter in the observed RM due to multi-path propagation effects (e.g., \citealt{Beniamini+22}).  It is not surprising that the RM fluctuations would be decoupled from any secular or periodic changes in the burst properties themselves (e.g., \citealt{Mckinven+22}), as the latter are produced within the relativistic jet on much smaller radial scales.  

For \prstwo~the DM is observed to decrease at the rate $-0.09 \pm 0.02$\,pc\,cm$^{-3}$\,day$^{-1}$ \citep{Niu+22}. From our models (Figs.~\ref{fig:Mdot1e5_nuLnu_RM_DM} and \ref{fig:FRB_fit}), we see that such a modest DM change is expected around the time the source transitions from free-expansion to self-similar expansion phase (i.e., $t\lesssim t_{\rm free}$). This is consistent with the estimated age of the system, $\sim10\,{\rm yr} \lesssim t_{\rm free}\sim13$\,yr.  This model further predicts that in the decades ahead, the $|$RM$|_{\rm max}$ and $|$DM$|_{\rm neb}$ will decrease at a slower rate and then flatten (mild increase) for the next $\sim10^3$\,yr (times $t>t_{\rm free}$ in the right panel of Fig.~\ref{fig:FRB_fit}).  However, we caveat that, other choices of the binary parameters could yield models that turns off quicker, but which exhibit otherwise similar emission properties at earlier times.  The late-time light curves and RM ($t\gg t_{\rm free}$) would differ from those predicted by our best-fit model (e.g., see right panel of Fig.~\ref{fig:Mdot_lc_RM_index}). 

The young source ages and resulting small PRS nebula radii we predict (Eq.~\ref{eq:R_n}) are also consistent with constraints on their sizes (e.g. $<0.6\,{\rm pc}$ from VLBI measurements of \prsone; \citealt{Marcote+17}). These contrast with the much larger ${\cal O}(100)$\,pc nebul\ae\ that surround many ULX in the local universe, due to their typically much greater ages $\sim10^6$\,yr.  This difference can potentially be understood as a selection effect: the most active and luminous FRBs (e.g., \prsone~and \prstwo\.) may preferentially arise from the highest $\dot{M}$, shortest-lived accretion phases ($\sim10^2$\,yr), because their larger jet powers $L_{\rm j} \propto \dot{M}$ are required to generate these most luminous FRBs.  The luminosity function of FRBs\footnote{The FRB luminosity function (top left panel of Fig.~\ref{fig:timescales}) is calculated using the burst flux measurements reported in the first CHIME/FRB catalog \citep{CHIME_catalog_21}. The distance to the sources are calculated assuming the NE2001 Galactic distribution of free electrons \citep{Cordes_Lazio_02}, and a fiducial FRB host galaxy DM of 50\,pc\,cm$^{-3}$ \citep{Luo+18,Luo+20}.} (top-left panel of Fig.~\ref{fig:timescales}) suggests that most of the FRBs---with $L_{\rm FRB}\gtrsim10^{41}$\,erg\,s$^{-1}$---require a mass transfer rate $\dot{M}/\dot{M}_{\rm Edd}\gtrsim10^4$.
On the other hand, lower-luminosity FRBs that can still be powered by the stable (thermal timescale) mass transfer by post-MS stars---which can have lifetimes of $\gtrsim 10^4$\,yr \citep{Klencki+21}---could exhibit large, ${\cal O}(100)$\,pc nebul\ae\ around them, closer to known ULX in the local universe.

Additional constraints on the ULX model for FRBs comes from the putative X-ray emission (Sec.~\ref{subsubsec:X-ray}).  For the high $\dot{M}$ systems needed to power the high radio luminosities of the observed PRS, the FS does not enter the radiative regime within the active timescale of the system, thus precluding X-ray emission commensurate with their high wind kinetic luminosities (Fig.~\ref{fig:L_X}).  On the other hand, putative lower-$\dot{M}$ systems which generate less luminous FRBs (with $L_{\rm FRB}\lesssim10^{41}$\,erg\,s$^{-1}$, or equivalently, $\dot{M}/\dot{M}_{\rm Edd}\lesssim10^4$) enter their radiative phase within their rather longer active lifetimes, generating more luminous shock emission $L_{\rm X} \gtrsim 10^{39}$ erg s$^{-1}$ (though still challenging to detect with current optical or X-ray telescopes at the $\gtrsim$ 100 Mpc distances of FRB PRS).  The existing X-ray non-detections of FRB sources are thus consistent with the X-ray binary model \citep{Sridhar+21b}.

Based on a search of non-nuclear FIRST radio sources in the local universe $<100$\,Mpc with a flux $\gtrsim 10\%$ that of the \prsone~PRS, and assuming a source lifetime of $10^{2}-10^{3}$\,yr, \cite{Ofek_17} estimate a birth rate of ${\cal R} <50-500\,{\rm yr}^{-1}\,{\rm Gpc}^{-3}$.  For this formation rate and our preferred PRS model with $v_{\rm j} \approx 0.2$, interpolating between the source count predictions in Fig.~\ref{fig:Mdot_detectability} shows that the predicted all-sky rate of PRS-like hyper-nebul\ae~would fall within the constraints imposed by VLASS.  

\subsection{Application to weaker, local ULXs} \label{subsec:local_ULX}
Our model can in principle also be applied to predict emission from lower-$\dot{M}$ ULX-bubbles in the local universe.  For example, ULX-1 in M51 exhibits non-thermal radio lobes with a 5 GHz luminosity $\nu L_{\nu} \approx 2\times 10^{34}$ erg s$^{-1}$, in comparison to the estimated total mechanical power of $L_{\rm w} \simeq 2\times 10^{39}$ erg s$^{-1}$ 
\citep{Urquhart+18}.  Other jetted ULX-bubbles exhibit similar radio luminosities within a factor of a few, e.g. Holmberg II X-1 \citep{Cseh+14,Cseh+15}, NGC 5408 X-1 \citep{Cseh+12,Soria+06,Lang+07}.  The resulting efficiency $\nu L_{\nu}/L_{\rm w} \sim 10^{-5}$ would appear to be substantially lower than the maximum radiative efficiency predicted by our model, $\sim \eta \varepsilon_{\rm e}(v_{\rm j}/c)^2 \gtrsim 10^{-3}$, in the fast-cooling limit.  However, the observed power-law synchrotron spectrum in ULX-1 is that of a {\it non-thermal} distribution electrons (for old, lower-$\dot{M}$ systems the thermal spectral peak is well below the GHz band; Eq.~\ref{eq:nusynth}), which likely receive a much smaller fraction of the jet power than the thermal electrons.  With modifications to include a non-thermal tail on the electron distribution injected into the nebula, our model could in principle be applied to these systems in future work.

Lower-$\dot{M}$ (possibly non-ULX) X-ray binaries could in principle also contribute to generating less luminous FRB observable in the local universe, for instance such as the the repeating FRB 20200120E located in a globular cluster in M81 \citep{Kirsten+22}.  The resulting lower power jets would generate less luminous radio synchrotron emission (see \citealt{Sridhar+21b} for more details), such as the persistent radio flux observed from Galactic X-ray binaries, whose luminosity $\nu L_{\nu} \sim 10^{30}$ erg s$^{-1}$ (e.g., \citealt{Fender&Hendry00}) is marginally consistent with constraints on the persistent radio flux from the location of FRB 20200120E ($\nu L_{\nu} \lesssim 4\times 10^{30}$ erg s$^{-1}$ at 1.5 GHz; \citealt{Kirsten+22}).

\section{Conclusion} \label{sec:conclusion}

We have developed a time-dependent one-zone model for the synchrotron emission from ``hyper-nebul\ae'' around ULX-like binaries. We dub them hyper-nebul\ae\ as they are inflated by the highly super-Eddington accretion disk/jet winds that accompany rapid runaway mass-transfer (typically $\dot{M}/\dot{M}_{\rm Edd}\gtrsim10^3$) from an evolved companion star (e.g., while crossing the Hertzsprung gap to become a giant star) onto a compact object (BH or NS), which can manifest at the final stages leading up to a common envelope event.  For concreteness, all of our calculations consider a fiducial binary containing a massive star ($M_\star=30M_\odot$) feeding a BH companion of fixed mass ($M_\bullet=10M_\odot$), and vary the active period of the system $t_{\rm active}$ by changing the mass accretion rate, $\dot{M}/\dot{M}_{\rm Edd} \in [10,10^7]$ (Sec.~\ref{subsec:binary}). The evolution of the intrinsic properties of the hyper-nebula and the observables are continuously evaluated across the free expansion ($t<t_{\rm free}$), deceleration ($t>t_{\rm free}$), and post-active ($t>t_{\rm active}$) stages (Sec.~\ref{subsec:disk_wind}).  This constant-$\dot{M}$ set-up is admittedly an idealization valid over a limited period of time, insofar that actual binaries will experience a range of mass-transfer rates over time, for instance with $\dot{M}(t)$ increasing leading up to the final dynamical plunge of the accretor into the donor's envelope (e.g., \citealt{Pejcha+17}).

Slow wide-angle disk winds dominate the total mass loss, and inflate the hyper-nebula equatorially, while faster (and potentially precessing) winds/jet from the innermost accretion flow are responsible for imparting a bipolar geometry to the nebula, and for energizing and injecting relativistic thermal electrons into the hyper-nebula (Fig.~\ref{fig:cartoon}).  We self-consistently evolve the hyper-nebula's size, magnetic field strength, and the injected electrons' energy distribution as a function of time---subject to various radiative and adiabatic losses (Sec.~\ref{subsubsec:electron_evolution}).  With this, we numerically evaluate the time evolution of specific observable properties such as the thermal synchrotron luminosity, rotation measure (and its fluctuation timescale), dispersion measure (through the shell and the nebula), and the energy spectrum, as a function of the wind ($\dot{M} = \dot{M}_{\rm w}$, $v_{\rm w}$), and jet ($\eta \equiv \dot{M}_{\rm j}/\dot{M}$, $v_{\rm j}$, $\sigma_{\rm j}$, $\varepsilon_{\rm e}$) properties (Sec.~\ref{sec:results}).

Our findings on the prospects of detecting hyper-nebul\ae\ are summarized as follows.  For fiducial choices of parameters, the disk wind-CSM forward shock enters radiative regime and shines bright in X-rays (primarily due to free-free emission) only for those systems accreting at rates $\dot{M}/\dot{M}_{\rm Edd}\lesssim10^4$, at times $t \gg t_{\rm free}$.  Even for these cases, the expected X-ray luminosity ($\sim10^{39-40}$\,erg\,s$^{-1}$) is challenging to detect at characteristic source distances $\gtrsim 100$ Mpc with current X-ray facilities, in part because they peak at soft temperatures $\sim0.1$\,keV (Sec.~\ref{subsubsec:X-ray}).  The accompanying optical counterpart (due to line-emission from shocked gas) may be detected by future surveys such as Rubin Observatory up to $\sim$100\,Mpc, or through HST follow-up observations of the closest radio-identified candidates.  Sensitive high-resolution follow-up observations (e.g., with the next generation VLA; \citealt{Murphy+18}) could in principle discern the `aspect ratio' of the hyper-nebul\ae\, thus placing constraints on the disk wind vs. jet velocities.

The prospects are better for detecting ULX hyper-nebul\ae\ as off-nuclear point sources by radio surveys such as VLASS or DSA-2000 \citep{Hallinan+19}. Although the detectable source population depends sensitively on the assumed system parameters (particularly those which control the electron heating), we estimate that up to $\sim10^3-10^6$ hyper-nebul\ae\ could exist within VLASS, among which ${\cal O}$(10) evolve sufficiently rapidly on timescales of years to decades to be identifiable as `transients' (Sec.~\ref{subsec:survey_detection}). 
These transients may presage energetic common envelope transients, such as LRNe or FBOTs, in the decades to centuries ahead, providing new constraints on the earliest stages of unstable mass-transfer in these candidate future gravitational wave sources detectable by the LIGO-VIRGO-Kagra collaboration \citep{LVC+18}; for the same reason, we encourage archival radio searches at the locations of these optical transient classes.  A comprehensive multi-wavelength study of the detection prospects (including e.g., hyper-nebul\ae\ size distribution, distance distribution, scintillation constraints, etc.) is deferred to future work.   

If a population of repeating fast radio bursts are powered by hyper-accreting compact objects \citep{Sridhar+21b}, the persistent radio source (PRS) spatially coincident with some of them could be explained as young ULX hyper-nebul\ae~(Sec.~\ref{subsec:frb_application}).  Unlike most ULX in the nearby universe, particularly young sources with ages $\lesssim 100$ yr, may be favored because their higher accretion powers may be requisite to powering the most luminous FRBs.  We demonstrate---as a proof of principle---that the radio spectrum of the PRS associated with \prsone~and \prstwo~are reproduced from the hyper-nebula model for close-to-fiducial assumptions.  Despite the narrow synchrotron spectrum predicted by our single-electron temperature baseline model, the broader observed spectra of the FRB PRS requires the electrons injected into the nebula possess a modest range of temperatures (corresponding to a $\lesssim 2-3$ temporal or spatial variation in the jet velocity).  The same model consistently accounts for the magnitude and time evolution of the $|$RM$|_{\rm max}$ of \prsone~and the characteristic timescales (though not the detailed evolution) of the RM variations about this secular trend (e.g., in \prsone, 20190520B, and 20180916B) due to turbulence within, or variations in the FRB line of sight through, the nebula.  

Aside from the extreme, high$-\dot{M}$ short-lived mass-transfer phases observable at cosmological distances that we have focused on here, a modified version of our model could be applied to the more volumetrically-abundant lower-$\dot{M}$ ULX-nebul\ae\ seen in the local universe (up to few tens of Mpc).  Although the models presented in this paper include only the emission from thermal electrons heated at the jet termination shock, diffusive shock acceleration or magnetic reconnection at the shock or in the nebula could also supply a population of non-thermal electrons.  This would generate the canonical power-law synchrotron spectrum extending to frequencies well above the thermal synchrotron peak, consistent with the radio spectra observed from radio-loud ULXs such as ULX-1, Holmberg II X-1 or NGC 5408 X-1.

\section{Acknowledgement} \label{sec:acknowledgement}
This paper benefited from useful discussions with Dillon Dong, Rui Luo, Ben Margalit, and Lorenzo Sironi. We thank Xian Zhang and Casey Law for sharing the FAST data for \prstwo.  N.S. acknowledges support from NASA (grant number 80NSSC22K0332).  B.D.M. acknowledges support from NASA (grant number 80NSSC22K0807).  The Flatiron Institute is supported by the Simons Foundation.

\appendix
A list of all the relevant timescales of the model introduced in this paper, and their definitions, are provided in Table~\ref{tab:timescales}.

\begin{deluxetable*}{ccl} \label{tab:timescales}
\tablenum{1}
\tablecaption{Model timescales.}
\tablewidth{0pt}
\tablehead{
\colhead{Variable} & \colhead{Cf.} & \colhead{Definition}
}
\startdata
$t_{\rm active}$ & Eq.~\ref{eq:t_active} & Total duration of the accretion phase at mass-transfer rate $\dot{M}$.\\
$t_{\rm th}$ & Eq.~\ref{eq:t_active} & Thermal timescale of the donor star losing mass to the accretor.\\
$P$ & Eq.~\ref{eq:t_active_orbperiod} & Orbital period of the binary.\\
$t_{\rm free}$ & Eq.~\ref{eq:tfree} & Free-expansion timescale beyond which the wind ejecta shell starts decelerating.\\
$t_{\rm ff}^{\rm thin}$ & Eq.~\ref{eq:t_ffthin} & Time after which the wind ejecta shell becomes optically thin to free-free emission.\\
$t_{\rm fs}^{\rm cool}$ & Eq.~\ref{eq:t_fs_cool} & Cooling timescale of the gas heated at the forward shock.\\
$t_{\rm exp}$ & Eq.~\ref{eq:ffcool_exp_timescale} & Expansion timescale of the shocked gas.\\
$t_{\rm cr}^{\rm fs}$ & Eq.~\ref{eq:t_fs_cr} & Critical time after which the forward shock becomes radiative.\\
$t_{\rm rs}^{\rm cool}$ & Eq.~\ref{eq:t_rs_cool} & Cooling timescale of the gas heated at the reverse shock.\\
$t_{\rm cool}^{\rm syn}$ & Eq.~\ref{eq:t_cool_syn} & Synchrotron cooling time of thermal electrons in the nebula.\\
$t_{\rm cr}^{\rm syn}$ & Eq.~\ref{eq:t_syn_cr} & Critical time after which nebular electrons radiate efficiently via synchrotron emission.\\
$t_{\delta {\rm RM}}$ & Eq.~\ref{eq:t_deltaRM} & Characteristic timescale for fluctuations in RM.\\
$t_{\rm pk}$ & Sec.~\ref{subsubsec:observable_properties}, Eq.~\ref{eq:N_tot} & Time when the peak radio luminosity is attained.\\
$\Delta t_{\rm pk}$ & Sec.~\ref{subsubsec:observable_properties}, Eq.~\ref{eq:N_tot} & Duration of peak light, defined as the time span over which the luminosity is $>0.5\times$ its peak value.\\
$\Delta t_{\rm i}$ & Eq.~\ref{eq:N_transient} & Time-interval over which the radio luminosity changes by a factor of $\gtrsim 3\times$ in 20\,yr.\\
\enddata
\end{deluxetable*}

\bibliographystyle{aasjournal}
\bibliography{ApJ}

\end{document}